\documentclass[lineno2]{jfm}

\usepackage{graphicx}
\usepackage{newtxtext}
\usepackage{newtxmath}
\usepackage{natbib}
\usepackage{siunitx}
\usepackage{tikz}
\usepackage{multirow}
\usepackage{comment}
\usepackage{hyperref}
\hypersetup{
    colorlinks = true,
    urlcolor   = blue,
    citecolor  = black,
}

\newcommand{\RomanNumeralCaps}[1]
\linenumbers

\usepackage{xcolor}

\newcommand{\DNSns}{\hat\beta_{\textit{NS}}(z)}  
\newcommand{\DNSs}{\hat\beta_{\textit{S}}(z)}    
\newcommand{\Expint}{\beta^i(z)}                     
\newcommand{\Expz}{\beta^0(z)}                     
\newcommand{\Expm}{\langle\beta^0\rangle}          
\newcommand{\DNSms}{\langle\hat\beta_S\rangle}          
\newcommand{\DNSmns}{\langle\hat\beta_{\textit{NS}}\rangle}          
\newcommand{\DNSeffs}{\hat\gamma_{\textit{S}}}              
\newcommand{\DNSeffns}{\hat\gamma_{\textit{NS}}}              

\title{Local slip length and surfactant effects on liquid-infused surfaces}

\author{S. Saoncella,
  J. Cerutti,
  T. Lenavetier,
  K. Amini,
  F. Lundell,
 \and S. Bagheri
\corresp{\email{shervinb@kth.se}}}
\affiliation{\aff{1}FLOW, Dept. Engineering Mechanics, Royal Institute of Technology  (KTH),  100 44 Stockholm, Sweden}

\begin{document}
\maketitle

\begin{abstract}
Robust surfaces capable of reducing flow drag, controlling heat and mass transfer, and resisting fouling in fluid flows are important for various applications. In this context, textured surfaces impregnated with a liquid lubricant show promise due to their ability to sustain a liquid-liquid layer that induces slippage. However, theoretical and numerical studies suggest that the slippage can be compromised by surfactants in the overlying fluid, which contaminate the liquid-liquid interface and generate Marangoni stresses. In this study, we use Doppler-optical coherence tomography, an interferometric imaging technique, combined with numerical simulations to investigate how surfactants influence the slip length of lubricant-infused surfaces with longitudinal grooves in a laminar flow. 
We introduce surfactants by adding tracer particles (milk) to the working fluid (water). Local measurements of slip length at the liquid-liquid interface are significantly smaller than theoretical predictions for clean interfaces \citep{Schonecker2013LongitudinalFluid}. In contrast,  measurements are in good agreement with numerical simulations of fully immobilized interfaces, indicating that milk particles adsorbed at the interface are responsible for the reduction in slippage. This work provides the first experimental evidence that liquid-liquid interfaces within textured surfaces can become immobilized in the presence of surfactants and flow.
\end{abstract}



\section{Introduction}
Lubricant-infused surfaces (LISs) submerged in liquid flows hold significant potential for drag reduction \citep{Solomon2014DragFlow,Rosenberg2016TurbulentSurfaces}, biofouling resistance \citep{Epstein2012Liquid-infusedPerformance}, and enhanced heat transfer \citep{Sundin2022HeatFlows}. These surfaces are characterized by a structured solid texture infused with a lubricant that is immiscible with the surrounding fluid. LISs are capable of self-healing and are not susceptible to failure induced by hydrostatic pressure. Therefore, LISs offer greater robustness than superhydrophobic surfaces (SHSs) in submerged environments when designed appropriately \citep{Wong2011BioinspiredOmniphobicity, Preston2017, Wexler2015Shear-drivenSurfaces, Sundin2021}.

The beneficial properties of LISs rest upon the existence of a stable and mobile liquid-liquid interface, which contribute to a slip effect on the external flow. The main factors that can impede the efficacy of LISs are lubricant drainage and the immobolization of the interface. 
The former is a recognised issue -- initially identified by \cite{Wexler2015Shear-drivenSurfaces} -- caused by prolonged exposure of a LIS to shear flow. Viscous stresses can deform the fluid interface, displacing the lubricant. For a LIS with a pattern of grooves parallel to the flow direction, a potential solution to prevent lubricant drainage is to confine the lubricant in cavities or segments \citep{Fu2019ExperimentalFlow} which are shorter than the maximum retention length prescribed by \cite{Wexler2015Shear-drivenSurfaces}. An additional retention mechanism has been demonstrated by \cite{saoncella2024contact}, which leverages the high contact angle hysteresis between the solid and the lubricant to stabilise lubricant droplets within the grooves when subjected to shear flow.

The second limiting factor (interface immobilization) can be induced by surface-active agents (surfactants) adsorbed at the liquid-liquid interface. While this phenomenon has been studied for SHSs, investigations of LISs are scarce. The investigations conducted for SHSs have demonstrated that the presence of surfactants at the gas-liquid interface can significantly reduce the slippage. The reduced slip on SHSs is caused by small gradients in concentration of surfactants adsorbed at the interface, which generate a  Marangoni stress opposing the flow \citep{Bolognesi2014EvidenceMicrochannel, Peaudecerf2017TracesSurfaces,Song2018EffectInterface}. Theoretical and numerical models support these observations and provide additional insights into the underlying mechanisms \citep{Landel2019ASurfactant, Temprano-Coleto2023AReduction, Baier2023InfluenceGrooves}. 

A recent numerical work by \cite{Sundin2022SlipSurfactants} used an adsorption model for alkane-water interfaces \citep{Fainerman2019} to investigate effects of surfactants on LISs consisting of transverse grooves and subjected to a shear flow. Based in the surfactant theory for SHSs (see e.g. \cite{Landel2019ASurfactant}), the authors developed an analytical model that predicts the effective slip length of the LIS in relation to the concentration of surfactants in the working fluid. Their findings demonstrate that the Marangoni stress significantly reduces the slip length, and indicate that the liquid-liquid interfaces of LIS are in fact more susceptible to slip reduction from surfactants compared to the liquid-gas interfaces of SHSs.
These results, highlighting the sensitivity of the liquid-liquid interface to contaminants, reinforce the necessity for experimental investigations, which are currently lacking.

\begin{figure}
    \tikzset{every picture/.style={line width=0.75pt}} 
    \raisebox{-1\height}{\begin{tikzpicture}[x=0.75pt,y=0.75pt,yscale=-1,xscale=1]
    \node [inner sep=0pt,below right,xshift=0.0\textwidth] 
                {\includegraphics[width=0.55\textwidth]{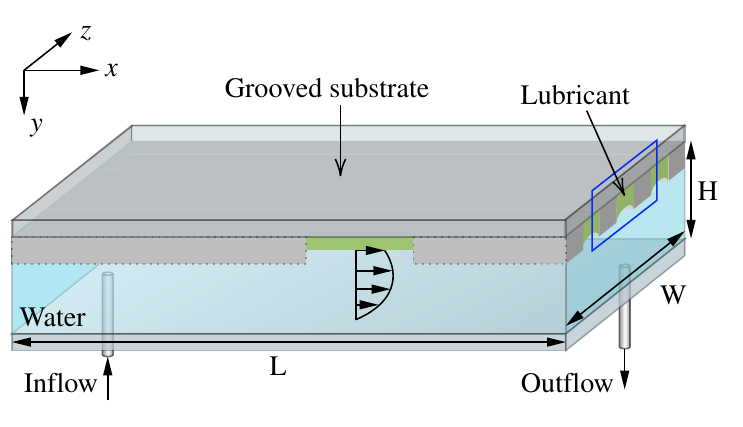}};
    \draw (-10,0) node [anchor=north west][inner sep=0.75pt]   [align=left] {($a$)};
    \end{tikzpicture}}
    \hfill
    %
    \tikzset{every picture/.style={line width=0.75pt}} 
    \raisebox{-1.3\height}{\begin{tikzpicture}[x=0.75pt,y=0.75pt,yscale=-1,xscale=1]
    \node [inner sep=0pt,below right,xshift=0.0\textwidth] 
                {\includegraphics[width=0.4\textwidth]{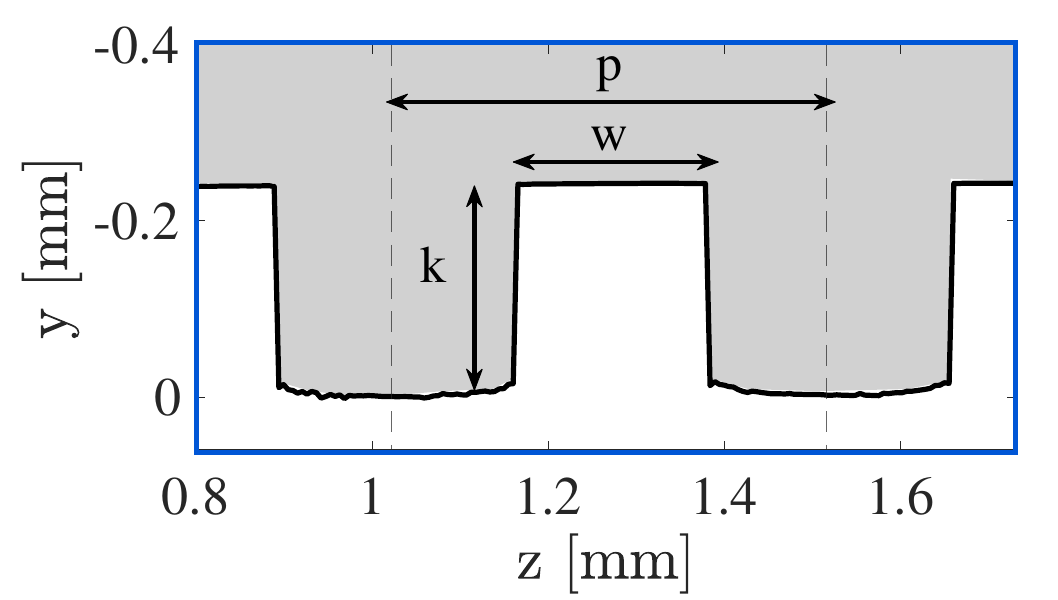}};
    \draw (-10,0) node [anchor=north west][inner sep=0.75pt]   [align=left] {($b$)};
    \end{tikzpicture}}
    
    \caption{($a$) Schematic representation of the experimental setup. The proportions of the sketch are not on a real scale. ($b$) Profile (measured using OCT) of the cross-section of a solid substrate used for the LIS. 
    The dimensions represented are the groove width, $w$, the groove depth, $k$, and the the pitch $p$.}
    \label{fig:setup}
\end{figure}

It is recognised that surface-active contaminants represents a significant challenge in  experimental investigations involving interfacial flows and surfaces.  For example, PDMS -- a common material used to fabricate microfluidic channels -- releases uncrosslinked oligomer chains \citep{Wong2020AdaptivePolydimethylsiloxane}, which are surface-active \citep{Hourlier-Fargette2017RoleElastomers} and diffuse into the  solution \citep{Carter2020PDMSApplications,Regehr2009BiologicalCulture}. \cite{Peaudecerf2017TracesSurfaces} and \cite{Sundin2022SlipSurfactants} report that trace amounts of surfactants in the bulk is sufficient to reduce the slip length, e.g.  bulk concentration of \qty{e-4}{\mol\per\meter\cubed} for SHS and \qty{e-5}{\mol\per\meter\cubed} for LIS results in a reduction with 50\%.

This study aims to quantify the slip length of LISs with longitudinal grooves to better understand the effects of surfactants. Doppler-optical coherence tomography (D-OCT), an interferometric imaging technique, is employed to measure the local velocity profile in a laminar duct with a LIS mounted on one wall. The OCT device is also used to extract the interface shape.  By complementing the experiments with targeted numerical simulations, we show that surfactants can nearly fully rigidify the liquid-liquid interface. 

The structure of this paper is as follows. Section \ref{sec:methods} describes the flow configuration and the experimental methodology. Section \ref{sec:results} presents the experimental results, comparing the local slip length to analytical models of LISs with and without surfactants. Section \ref{sec:numerics} discusses the numerical results and their comparison to experiments, providing indirect evidence that surfactant contamination is responsible for the reduced slippage observed experimentally. Finally, Section \ref{sec:conclusion} summarizes the paper and discusses its main outcomes.

\section{Configuration and methods}
\label{sec:methods}
This section presents the flow configuration and the design of LISs used in the experiments. The methodology and the scheme used to measure the shape of the liquid-liquid interface and the flow velocity are also described.

\subsection{Flow cell and LIS design}
%
Figure \ref{fig:setup}$a$ shows the flow channel used in this study, with height $H=\SI{1}{\milli\meter}$, width $W=\SI{3.25}{\milli\meter}$ and length $L=\SI{80}{\milli\meter}$. The top surface is patterned with rectangular grooves parallel to the flow direction, which constitute the solid substrate of the LIS.
The inlet and outlet of the flow are located at the extremities of the bottom wall through holes of $\SI{1}{\milli\meter}$ in diameter.
The detailed description of the channel assembly is reported in Appendix \ref{app:A1}.
Hexadecane (Sigma-Aldrich), with dynamic viscosity $\mu_l=\SI{2.7}{\milli\pascal\per\second}$, is employed as a lubricant to impregnate the grooves, while the working fluid consists of a solution of water and milk at \SI{20}{\percent} concentration (later referred to as simply water) with dynamic viscosity $\mu_w$. The viscosity ratio between water and lubricant is defined as $N=\mu_w/\mu_l$. 

\begin{table}
  \begin{center}
\def~{\hphantom{0}}
  \begin{tabular}{lccccccc}
    Case & $w$ $[\SI{}{\micro\meter}]$ &  $k$ $[\SI{}{\micro\meter}]$ & $p$ $[\SI{}{\micro\meter}]$ & $A$ & $a$ & $N$ & $\phi$\\[3pt]
    A059N03 & 376 $\pm$ 5 & 223 $\pm$ 3 & 621 $\pm$ 2 & 0.59 & 0.61 & 0.3 & \SI{32.1}{\degree} $\pm$ \SI{1.2}{\degree}\\
    A089N03 & 271 $\pm$ 5 & 242 $\pm$ 3 & 608 $\pm$ 2 & 0.89 & 0.45 & 0.3 & \SI{20.6}{\degree} $\pm$ \SI{2.9}{\degree}\\ 
    A106N03 & 223 $\pm$ 5 & 237 $\pm$ 3 & 495 $\pm$ 2 & 1.06 & 0.45 & 0.3 & \SI{24.6}{\degree} $\pm$ \SI{2.3}{\degree}\\
    A106N1 & 223 $\pm$ 5 & 237 $\pm$ 3 & 495 $\pm$ 2 & 1.06 & 0.45 & 1.2 & \SI{21.2}{\degree} $\pm$ \SI{1.7}{\degree}\\
    A106N13 & 223 $\pm$ 5 & 237 $\pm$ 3 & 495 $\pm$ 2 & 1.06 & 0.45 & 13.2 & \SI{29.8}{\degree} $\pm$ \SI{2.9}{\degree}\\
  \end{tabular}
  \caption{The parameters defining the LISs for the five configurations: groove width $w$, groove depth $k$, groove pitch $p$, aspect ratio $A=k/w$, slipping surface fraction $a=w/p$, viscosity ratio $N=\mu_{w}/\mu_l$ and protrusion angle of the meniscus $\phi$.}
  \label{tab:LIS_cases}
  \end{center}
\end{table}

A total of five experimental cases with different geometrical features and viscosity ratio were designed (see Table \ref{tab:LIS_cases}).
Three grooved surfaces, differing by the groove aspect ratio, are fabricated in transparent resin using a casting technique (see Appendix \ref{app:A2}).
An example of a groove profile is shown in Figure \ref{fig:setup}$b$.
The grooves have depth $k$ of about \SI{230}{\micro\meter}, while their widths $w$ vary from 223 to \SI{376}{\micro\meter}, resulting in aspect ratios $A=k/w$ ranging from 1.06 to 0.59. 
The slipping area fraction, defined as $a=w/p$, ranges between 0.45 and 0.61. The influence of the viscosity ratio on the slip length was tested using the LIS with aspect ratio $A=1.06$. The water solution was mixed with glycerol in mass ratios of 1:0.7 and 1:3.5, increasing $N$ from $0.3$ to $1.2$ and $13.2$, respectively. As detailed in Appendix \ref{app:A2}, the LIS is prepared by first filling the duct with lubricant and then with the water solution at creeping flow. The lubricant remains trapped inside the grooves by capillary forces. After complete filling of the duct, the flow rate was gradually increased to \SI{0.4}{\milli\liter\per\minute}, which provides a bulk velocity $U_b=\SI{2}{\milli\meter\per\second}$ and a Reynolds number $\Rey\approx1$. 

\subsection{Optical coherence tomography}
\begin{figure}
    \centering
    \includegraphics[width=0.65\linewidth]{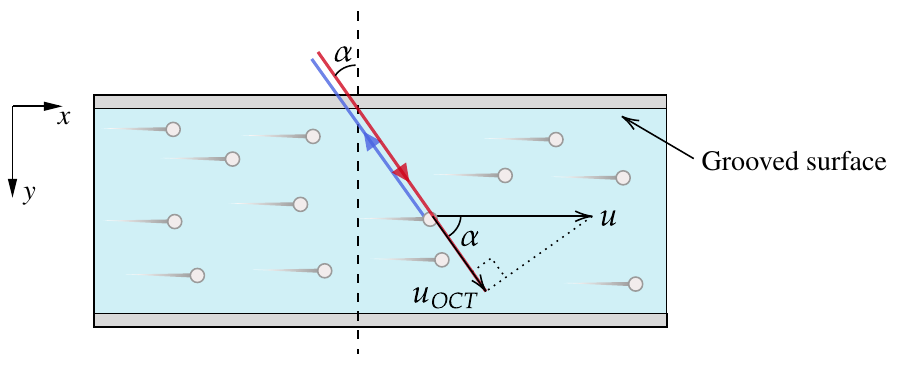}
    \caption{Schematic of velocity components obtained through D-OCT. The phase of the incoming reference beam (red) is compared with that of the backscattered beam (blue) from the moving particle to obtain the parallel component of the particle velocity.
    }
    \label{fig:OCT}
\end{figure}
Optical coherence tomography (OCT) is a low-coherence interferometry-based technique used to generate images that capture optical scattering within opaque samples on a millimeter scale, with a resolution of just a few micrometers.
While OCT has primarily been used in medical field \citep{Yonetsu2013OpticalCardiology},
recently it has been employed also in microfluidics and soft matter research \citep{Huisman2024NoninvasiveReview,Haavisto2015RheologicalTomography,Lauri2021OnlineTomography,KrishneGowda2019EffectiveExperiments,Gupta2024ShearGels,Lauga2005HandbookDynamics,amini2025optical}.

The functioning principle of OCT is based on a Michelson interferometer. A broadband light source is split into two components using a beam splitter. One component is directed into a reference arm, while the other is transmitted through the sample. The light reflected from the sample is recombined at the beam splitter with the reference beam, producing an interference signal. When the optical-path-length difference is within the coherence length
of the light source, interference occurs \citep{Tomlins2005TheoryTomography}. The output spectrum is analyzed by a spectrometer, generating a complete interference pattern that represents variations in intensity as a function of depth. Within this pattern, refractive index variations between layers in the sample manifest as intensity magnitude variations, providing detailed information about the sample's depth.

When there is a flow, the OCT imaging technique can be combined with the acquisition of a Doppler frequency (D-OCT) to obtain velocimetry information \citep{Chen1997NoninvasiveTomography}. The interference of light backscattered from a moving particle with the reference beam produces a beating at the Doppler frequency, which generates a corresponding Doppler phase shift. This phase shift is proportional to the projected component of the velocity measured along the OCT beam, $u_{OCT}$, as illustrated in Figure \ref{fig:OCT}. To reconstruct the velocity component in the $x$ (streamwise) direction,  an angle $\alpha$ between the beam and the vertical direction is required, yielding $u=u_{OCT}/\sin{\alpha}$. The velocity resolution depends on the acquisition time and the scan angle.
We used a Telesto II Spectral Domain OCT apparatus (Thorlabs Inc., NJ, USA), which has a central wavelength of \SI{1310}{\nano\meter} and a nominal bandwidth of \SI{270}{\nano\meter}, leading to a depth resolution of \SI{2.58}{\micro\meter} in water.

In these experiments, commercial milk is added to water at a volume ratio of 1:5 as a contrast medium.
No significant change in the viscosity, interfacial tension or refractive index of water is measured at this concentration of milk. As milk mixes with water rather than with the lubricant, the contrast in the intensity signal allows the two fluid components to be clearly distinguished.
Optical coherence tomography is used here to image the LIS inside the duct during flow and in Doppler mode for the acquisition of velocity profiles. The flow cell is positioned below the OCT objective at an angle $\alpha\approx\SI{6}{\degree}$. The OCT beam enters the duct through the clear top wall.

\subsubsection{Intensity measurements}
\begin{figure}
  \centerline{\includegraphics[width=1\textwidth]{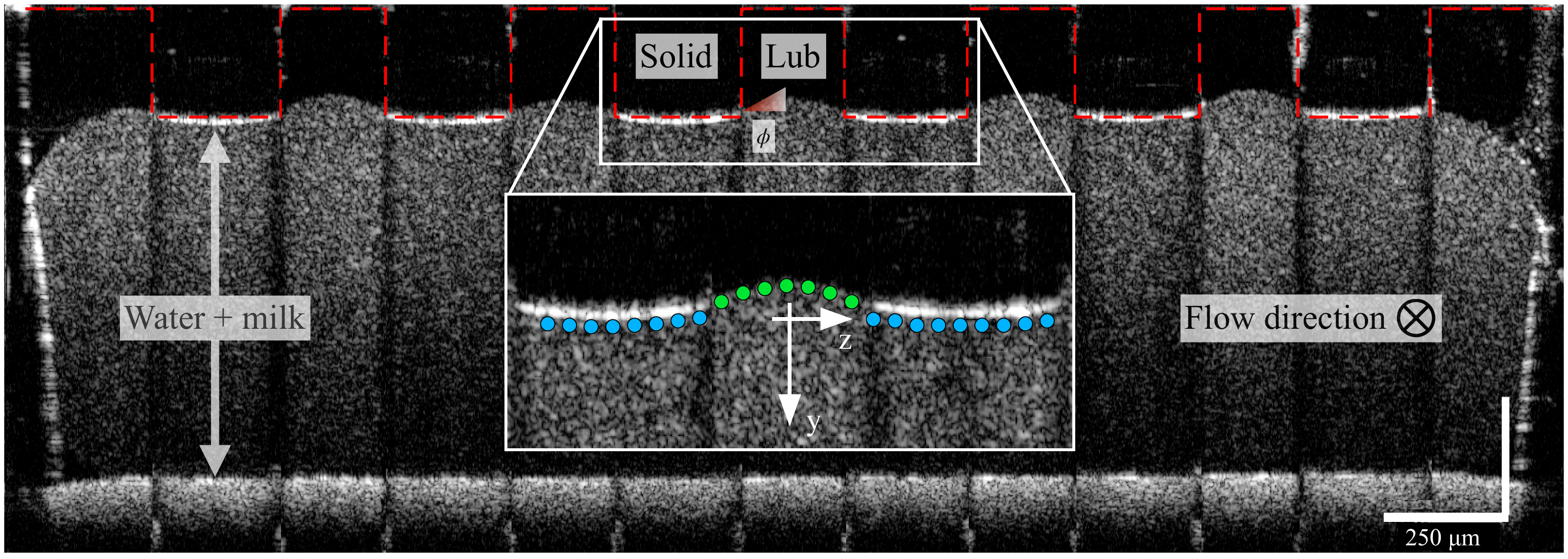}}
  \caption{Cross-sectional scan of the duct acquired with OCT. The grooves, marked as "Solid" are mounted on the top wall and their profile is highlighted with a red dashed line. The space between the grooves is filled with lubricant, marked with "Lub", whose profile bows at an angle $\phi$ to the top of the grooves. The volume of the duct is filled with a water-milk mixture which appears opaque. The insert shows an enlargement of the measurement positions marked with blue dots at the water-solid interface and with green dots at the water-lubricant interface.}
\label{fig:cross_sect}
\end{figure}

The intensity signal acquired with OCT across the flow cell allowed us to monitor the distribution of lubricant within the grooves during the initial infusion and the subsequent measurements. 
As an example, Figure \ref{fig:cross_sect} shows a cross-section of the flow cell during flow with the LIS mounted at the top wall (highlighted with a dashed line) and lubricant within the grooves. From this tomogram, it can be seen that the water-lubricant interface is curved with a meniscus bending towards the top. This is due to a pressure difference across the interface and to the hydrophobicity of the grooved wall. The streamwise average deflection $\phi$ of the liquid interfaces from the level $y=0$ and the associated standard deviation is derived from the tomograms and reported in Table \ref{tab:LIS_cases}.  

\subsubsection{Velocity measurements}

\begin{figure}
\centering
    \tikzset{every picture/.style={line width=0.75pt}} 
    \raisebox{-1\height}{\begin{tikzpicture}[x=0.75pt,y=0.75pt,yscale=-1,xscale=1]
    \node [inner sep=0pt,below right,xshift=0.0\textwidth] 
                {\includegraphics[width=0.46\textwidth]{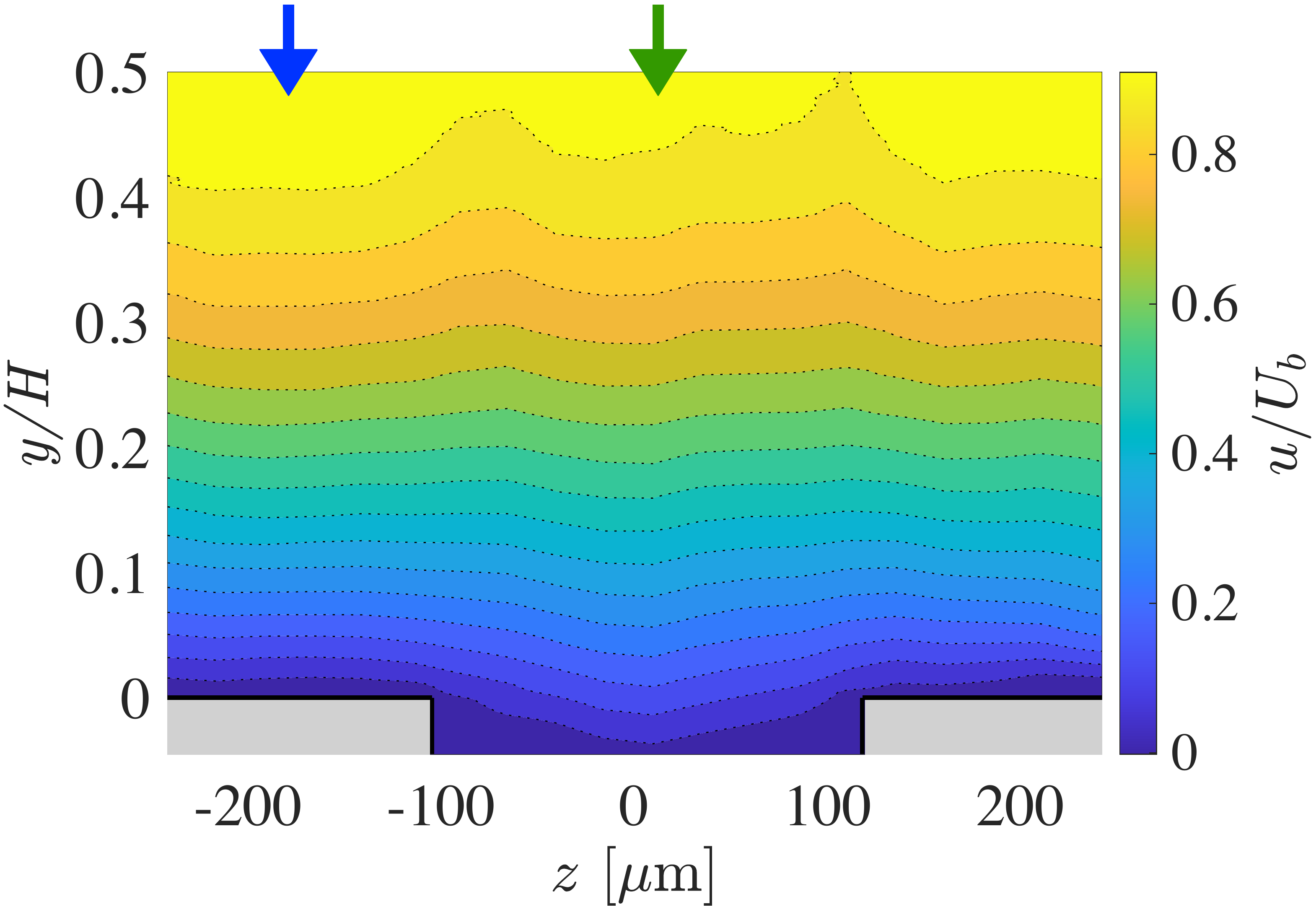}};
    \draw (-10,0) node [anchor=north west][inner sep=0.75pt]   [align=left] {($a$)};
    \end{tikzpicture}}
    \tikzset{every picture/.style={line width=0.75pt}} 
    \raisebox{-1\height}{\begin{tikzpicture}[x=0.75pt,y=0.75pt,yscale=-1,xscale=1]
    \node [inner sep=0pt,below right,xshift=0.0\textwidth] 
                {\includegraphics[width=0.46\textwidth, trim={0 0 0 0},clip]{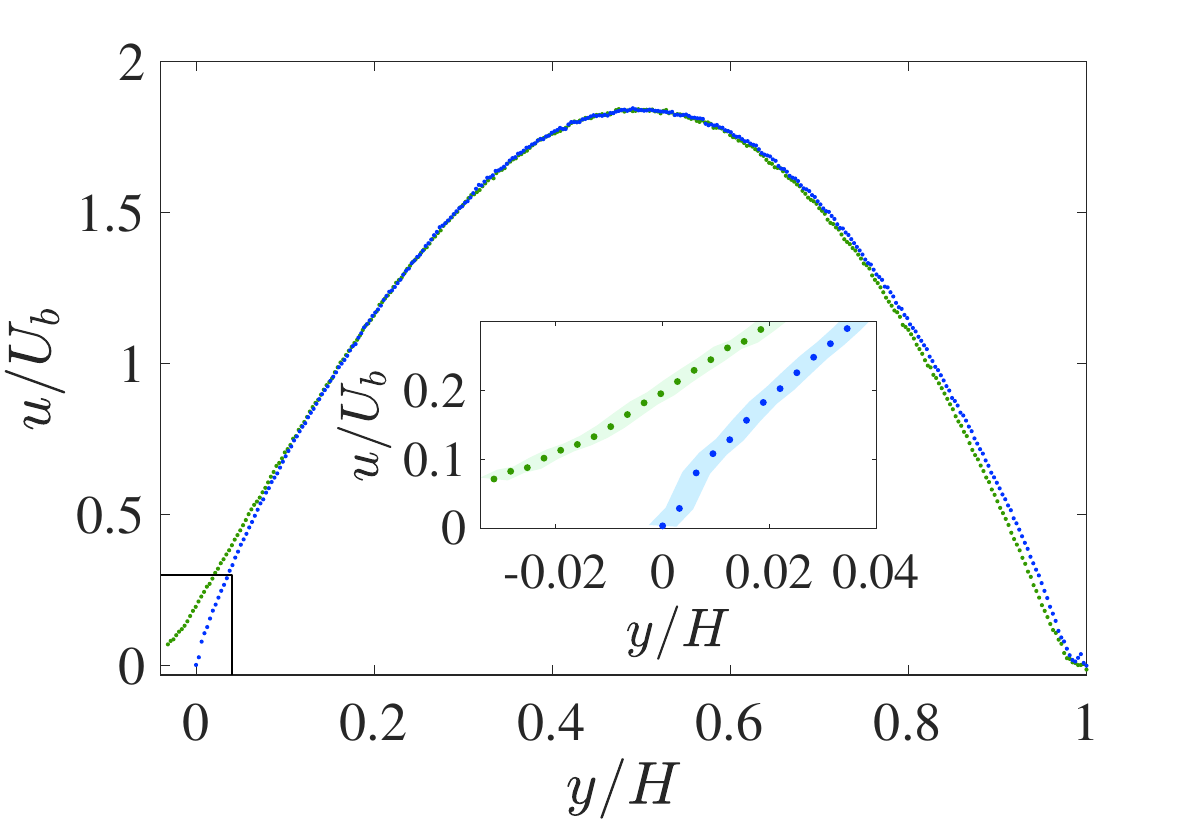}};
    \draw (-10,0) node [anchor=north west][inner sep=0.75pt]   [align=left] {($b$)};
    \end{tikzpicture}} 
    \caption{($a$) Contour plot of the velocity measurements in the central region of the duct for case A106N13. The arrows indicate the location of the measurements shown in frame ($b$). ($b$) Velocity profiles were acquired at the water-lubricant interface (green markers) and the water-solid interface (blue markers), for the same case. In the insert, the coloured area surrounding the markers represents the standard deviation of the velocity.}
    \label{fig:velocity}
\end{figure}

The measurements of velocity profiles were conducted through the full depth of the duct at the groove closest to the axial centre of the duct, where the influence of side walls on the flow is minimal. The acquisition frequency was \SI{5.5}{\kilo\hertz}. The locations of the local measurements are shown in the insert of Figure \ref{fig:cross_sect} where they are marked by blue-filled circles at the water-solid boundary and by green circles at the water-lubricant interface. The spanwise spacing between them is \SI{30}{\micro\meter}.
For each configuration, five spanwise series of velocity profiles straddling a pitch were recorded in streamwise locations spaced by 4.5, 5.0, 5.5, 6.0 and \SI{6.5}{\centi\meter} from the duct inlet, where the flow was fully developed. 

The velocities measured in the vicinity of the groove, normalised with the bulk velocity $U_b$, are shown as a contour plot in Figure \ref{fig:velocity}$a$ for case A106N13. Note that the y-axis is now directed upwards and the grooved surface is at the bottom. The black solid lines represent the solid ridges, while the blue and green arrows show the locations of the velocity profiles displayed in Figure \ref{fig:velocity}$b$.
The two velocity profiles in Figure \ref{fig:velocity}$b$ are shown using the same colour code as the arrows in Figure \ref{fig:velocity}$a$. The top and bottom walls of the flow cell have coordinates $y=0$ and $y=H$, respectively. The velocity profile measured on the solid ridge is analogous to a Poiseuille flow, exhibiting zero velocity at the walls. In contrast, the profile measured on the liquid-liquid interface deviates from the Poiseuille profile due to the presence of slip at the interface. The increased distance from the top wall of the green profile is due to the interface bending toward the groove bottom.
In the insert of Figure \ref{fig:velocity}$b$, an enlargement of the velocity profiles shows the uncertainty intervals of the measurements calculated as the standard deviation which, near the boundary, equals to $\pm\SI{5}{\percent}$ of the measured value.

\section{Slip of LIS}
\label{sec:results}
This section presents the experimental results and provides a comparative analysis with analytical predictions. The difference between the experiments and theory is discussed taking into account the influence of contaminants on the mobility of the liquid-liquid interface. 

\subsection{Local slip length}
\begin{figure}
  \centerline{\includegraphics[width=1\textwidth]{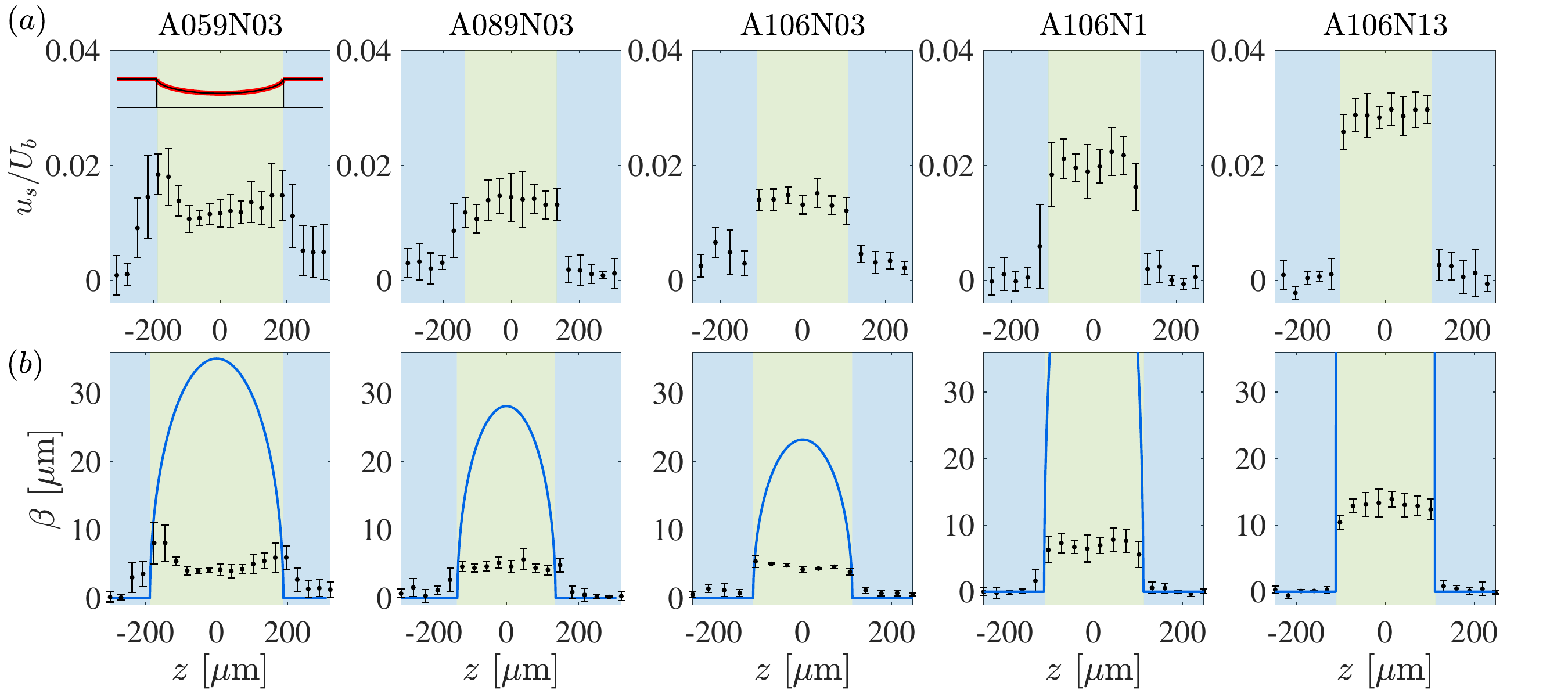}}
  \caption{($a$) Streamwise-averaged slip velocities as a function of the spanwise position $z$. The error bars correspond to the standard deviations. The sketch in the left panel indicates that the values are calculated at the interface. ($b$) Local slip lengths obtained from measurements, $\Expint$ (markers) and from the analytical model of \protect\cite{Schonecker2013LongitudinalFluid}, $\beta^{\textit{Sch}}(z)$ (solid line). In the fourth and fifth frames, the curve corresponding to the model exceeds the boundary of the plot. The background colour of the tiles identifies the region of the liquid-liquid interface (green) and the region of the liquid-solid interface (blue).}
\label{fig:localSL_SV}
\end{figure}

Figure \ref{fig:localSL_SV}$a$ shows the streamwise-averaged slip velocities $u_s(z)$ normalized by the bulk velocity $U_b$, measured along both the liquid-solid and liquid-liquid interfaces within a single pitch. The first three frames (from left to right) correspond to cases with increasing aspect ratio and constant viscosity ratio (A059N03, A089N03, and A106N03). The fourth and fifth frames represent configurations with the same geometry as the third column but with higher viscosity ratios (A106N1 and A106N13). The green background represents the physical width of the groove -- containing the liquid-liquid interface -- while the blue colour represents the area occupied by the liquid-solid interface. In all configurations, the slip velocity at the solid boundary is approximate $\SI{0.2}{\percent}$ of $U_b$. At the liquid interface, the slip velocity averages $\SI{1.3}{\percent}$ of $U_b$ for the cases with 
$N=0.3$, increasing to $\SI{2}{\percent}$ and $\SI{3}{\percent}$ for $N=1$ and $N=13$, respectively. 
The variance associated with the slip velocities are calculated as the standard deviation of repeated scans whose average gives a single velocity measurement.

\begin{table}
  \begin{center}
\def~{\hphantom{0}}
  \begin{tabular}{lcccccc}
    Case & $\beta^{\textit{Sch}}(0)$ $[\SI{}{\micro\meter}]$ & $\langle\beta^i\rangle_{lub}$ $[\SI{}{\micro\meter}]$ & $R$ & $\tau_l\cdot10^{-3}$ $[\unit{\pascal}]$ & $\tau_w$ $[\unit{\pascal}]$ & $\tau_l/\tau_w$ \\[3pt]
    A059N03 & 34.0 & 5.1 $\pm$ 1.5 & 6.8 &1.5 $\pm$ 0.2 &0.0057 $\pm$ 0.0001 & 0.26 $\pm$ 0.04\\
    A089N03 & 28.1 & 4.7 $\pm$  0.5 & 5.9 &1.4 $\pm$ 0.2 &0.0055 $\pm$ 0.0002 & 0.26 $\pm$ 0.04\\  
    A106N03 & 23.2 & 4.6 $\pm$ 0.5 & 5.0 & 1.6 $\pm$ 0.2&0.0052 $\pm$ 0.0002 & 0.31 $\pm$ 0.05\\
    A106N1 & 77.3 & 6.8 $\pm$ 0.7 & 11.3 & 2.2 $\pm$ 0.2&0.0186 $\pm$ 0.0003 & 0.12 $\pm$ 0.01\\
    A106N13 & 1004.9 & 12.7 $\pm$ 1.0 & 78.9 &3.0 $\pm$ 0.3 &0.1917 $\pm$ 0.0036 & 0.015 $\pm$ 0.002\\
  \end{tabular}
  \protect\caption{Central slip length value predicted by \cite{Schonecker2013LongitudinalFluid}, $\beta^{\textit{Sch}}(0)$, average slip length over the lubricated groove, $\langle\beta^i\rangle_{lub}$, ratio $R=\beta^{\textit{Sch}}(0)/\langle\beta^i\rangle_{lub}$, estimated shear stress in the lubricant $\tau_l$, measured shear stress at the water-lubricant interface $\tau_w$ and their ratio.}
  \label{tab:beta_comparison}
  \end{center}
\end{table}

For each velocity profile the ratio between the slip velocity and the velocity gradient $\partial u/\partial y$ is calculated and the local slip length at the interface, $\beta^i(z)$, is evaluated as the average of the repetitions at $n$ streamwise positions, ($x_j=4.5,5.0,5.5,6.0,\SI{6.5}{\centi\meter}$),
\begin{equation}
    \beta^i(z)=\dfrac{1}{n}\sum_{j=1}^n\frac{u_s(z,x_j)}{\frac{\partial u(z,x_j)}{\partial y}|_{\textit{interface}}}.
\label{eq:beta_loc_exp}
\end{equation}
The superscript $i$ denotes variables evaluated at the curved liquid-liquid interface, corresponding to the red curve in the sketch in the first panel of Figure \ref{fig:localSL_SV}$a$.
The variation in slip velocity between the solid and liquid boundaries translates directly to the corresponding slip length values shown in Figure \ref{fig:localSL_SV}$b$. 
At the liquid-solid interface, the measured slip length remains consistently around $1\pm\SI{1}{\micro\meter}$ across all cases. In contrast, the slip length at the liquid-liquid interface is $5\pm\SI{1}{\micro\meter}$ for the first three cases, increasing to $7\pm\SI{2}{\micro\meter}$ and $13\pm\SI{2}{\micro\meter}$ for the fourth and fifth cases, respectively. The uncertainty associated with the local slip length is the standard deviation of the repeated measurements in the different streamwise positions (see Appendix \ref{appB}).

To understand how the slip length of a realistic LIS setup differs from an ideal theoretical configuration, we compare our results to the local slip length predicted by the model of \cite{Schonecker2013LongitudinalFluid}, here named $\beta^{\textit{Sch}}(z)$. Their analytical equation characterizes the variation along the pitch of the local slip length of a shear flow over an infinitely long rectangular groove aligned with the flow direction. The cavity contains a second fluid with different viscosity, and the interface between the two fluids is flat. The local slip length is modelled as an elliptic function with a maximum value, $D$ at the centre of the groove (where the flow is least affected by the side walls of the cavity itself). This expression reads,
\begin{equation}
    \beta^{\textit{Sch}}(z)=\Real\Biggl\{-i2ND\sqrt{z^2-\frac{w^2}{4}}\Biggl\}.
\label{eq:schonecker}
\end{equation}
The modelling of $D$ is discussed in depth by \cite{Schonecker2013LongitudinalFluid} and is based on previous investigations of lid-driven cavity flows. Note that, in their consecutive work \citep{Schonecker2014InfluenceState}, the analytical expression of the local slip length distribution was derived for an array of parallel lubricant-filled grooves, which accounts for a modification of the velocity field when the distance between the grooves is small ($a\to1$). In our case, however, the simpler expression of single groove \eqref{eq:schonecker} is sufficiently accurate since the slip fraction is small (the difference between the expressions is smaller than \SI{2}{\percent}). 

Figure \ref{fig:localSL_SV}$b$ compares the experimental results with the analytical solution \eqref{eq:schonecker}.
The measured values deviate from the elliptical profile predicted by the model; we observe instead a flat shape with a nearly constant local slip along the interface. We also observe that the spanwise-average slip length measured at the interface, $\langle\beta^i\rangle_{lub}$, is significantly lower than the central value of the model distribution, $\beta^{\textit{Sch}}(0)$. Specifically, as reported in Table \ref{tab:beta_comparison}, the ratio $R$ between those two values varies from $5$ to $7$ for cases with small $N$ and increases to $80$ in the case with the highest viscosity ratio.

These observations may be explained by considering how our setup deviates from the ideal model. The three most evident differences are: (i) the finite boundaries of the duct which limit the number and the length of the grooves; (ii) the curvature of the water-lubricant interface; and (iii) the presence of the scattering medium in the water phase that can affect the dynamics of the interface. 

Regarding the first point, we can deduce from the velocity measurements that in the central region of the duct (see Figure \ref{fig:velocity}$a$), the spanwise velocity variations are small, as the velocity field is nearly uniform in this region. On the other hand, the finite length of the grooves can cause a backflow of the lubricant opposing the water flow, contributing to a reduction of slip velocity with respect to the case of infinitely long grooves. The contribution of the shear stress arising from this backflow is discussed in the next section.

\begin{figure}
    \centering
    \includegraphics[width=0.8\linewidth]{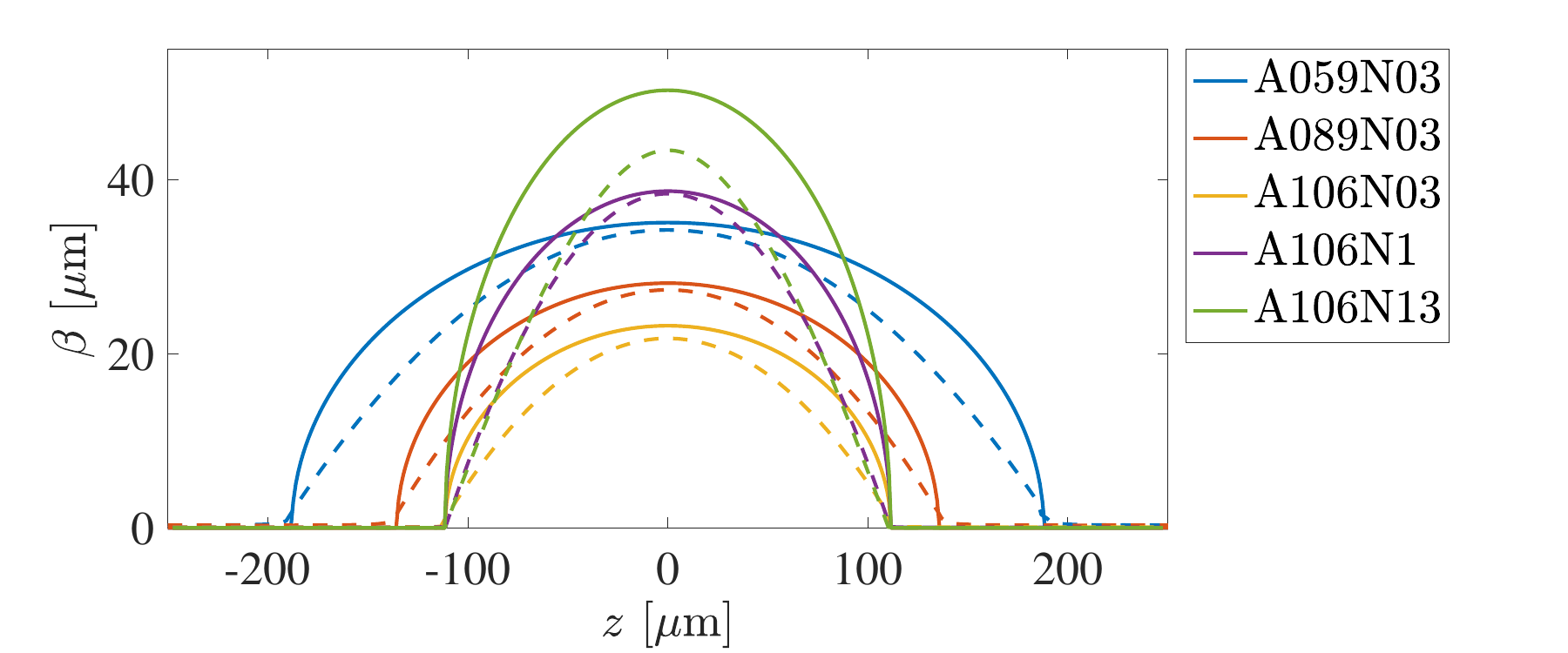}
    \caption{Effect of the curved interface on the local slip length. Continuous lines correspond to $\beta^{\textit{Sch}}(z)$, dashed lines to the results of DNS (with meniscus curvature according to Table \ref{tab:LIS_cases}). The curves representing case A106N1 are scaled by a factor of 2 and those representing case A106N13 are scaled by a factor of 20 for better visualisation.}
    \label{fig:meniscus}
\end{figure}

To investigate the impact of the meniscus at the liquid-liquid interface on the flow, numerical simulations were conducted. The computational setup replicated the LIS designs and enforced the interface curvature observed in the experiments (Table \ref{tab:LIS_cases}).
A detailed description of the numerical setup and methods is provided in Section \ref{sec:numerics}. As shown in Figure \ref{fig:meniscus}, the slip lengths obtained from simulations (dashed lines) at a curved interface are slightly smaller than the theoretical predictions of \cite{Schonecker2013LongitudinalFluid}. This indicates that the presence of a concave interface does not dramatically modify the slip length distribution as observed in Figure \ref{fig:localSL_SV}$a$.
The third point, i.e. the effect of contrast medium in the working fluid, is treated in the following section. 

\subsection{Surfactant contamination} \label{sec:surfactant}
As described in Section \ref{sec:methods}, to measure the flow velocity using D-OCT, the fluid must be light-scattering in the near-infrared. 
For this purpose, regular milk is mixed with water, as it provides a solid Doppler signal. However, milk might affect the interfacial properties between water and the lubricant infusing the LIS. 
Bovine milk contains two major proteins: casein and $\beta$-lactoglobulin, which account for approximately ~\SI{80}{\percent} and ~\SI{20}{\percent} of the total protein content, respectively \citep{OMahony2014Milk:Overview}. Both these proteins have a hydrophilic and an oleophilic part to their structure, as detailed in \cite{McSweeney2013AdvancedAspects}. We may therefore regard them as soluble surfactants, which can accumulate at the water-lubricant interface, as well as being in solution.
In a surfactant-laden fluid, the surfactant molecules adsorb at the interface (gas-liquid or liquid-liquid) and, advected by the flow, may accumulate at stagnation points \citep{Manikantan2020SurfactantFlows}. This effect leads to a gradient of surfactant concentration which generates Marangoni stresses opposing the flow.

The observation that the slip velocity at the interface is significantly lower than predicted by the analytical model can be explained by taking into account the Marangoni stress into the tangential stress balance at the interface. At the water-lubricant interface the wall-shear stress of the working fluid, $\tau_w$, is balanced by two contributions. The first is given by the viscous resistance of the lubricant inside the groove, $\tau_l$, while the second contribution is given by the Marangoni stress, $\tau_{\textit{Ma}}$. Thus, we can write,
\begin{equation}
    \tau_w = \tau_l + \tau_{\textit{Ma}}.
\end{equation}
The shear stress of the bulk fluid is calculated from the velocity measurements as $\tau_w=\mu_w {\partial u}/{\partial y}$ at $y=0$. 
The shear stress imposed by the lubricant on the interface can be estimated by assuming a velocity profile in the cavity. Neglecting interface curvature and confinement effects from the side walls of the duct, two limiting cases can be distinguished.
In one case, we may assume that the lubricant is not conserved, i.e. lubricant is allowed to leave the grooves at the very downstream position, and if no new lubricant is entering upstream, the groove will eventually be drained (albeit very slowly). In this case, we expect a Couette profile in the groove and $\tau_l = \mu_l {u_s}/{k}$, where $u_s$ is the slip velocity.
In the second case, the lubricant is conserved because of a physical barrier downstream that creates a recirculating flow in the groove. We then have a pressure gradient that opposes the Couette flow to enforce lubricant conservation across the cavity section and $\tau_l = 4 \mu_l {u_s}/{k}$ \citep{Busse2013ChangeSurface}.

An estimation of the Marangoni stress can be obtained by computing the ratio of $\tau_l$ to $\tau_w$. A stress ratio near unity means that $\tau_{\textit{Ma}}\ll\tau_l$, whereas a ratio near zero indicates $\tau_{\textit{Ma}}\gg\tau_l$. As shown in Table \ref{tab:beta_comparison}, Marangoni stresses dominate over the lubricant shear stress as $\tau_l/\tau_w \leq 0.3$ for all the configurations.
In particular, we find that $\tau_{\textit{Ma}}\approx0.70\tau_w$ for the cases at low viscosity ratios and increases to $0.88\tau_w$ and $0.98\tau_w$ for $N=1$ and $N=13$, respectively.

A related configuration was treated by \cite{Sundin2022SlipSurfactants}, who modelled a two-dimensional flow carrying surfactants over transverse lubricated grooves. Their model is adapted to the present configuration, which instead involves longitudinal flow, using the steps detailed in Appendix \ref{appC}.
The model derives the following expression of the effective slip length in the presence of surfactants in the bulk flow,
\begin{equation}
    \gamma^{\textit{Sun}}_l = b_{\textit{SHS,l}} \ \beta_{\textit{LIS,l}} \left( 1- \frac{1}{1+\alpha_{\textit{S,l}} + \alpha_{\textit{diff,l}}} \right).
\end{equation}
Here, $b_{SHS,l}$ is the equivalent slip length for a clean SHS with the same geometry, $\beta_{LIS}$ is a prefactor accounting for the non-zero viscosity in the grooves of a LIS and $\alpha_{S,l}$ and $\alpha_{\textit{diff,l}}$ are coefficients related to the transport processes of surfactants. The former accounts for the capability of surfactants to desorb as they are advected downstream, the latter accounts for the competition between advection and diffusion of surfactants at the interface.
The exact expressions of these terms as well as of the non-dimensional numbers involved are shown in Table \ref{tab:non-dim-numb} along with characteristic values.

The expression for $\gamma^{\textit{Sun}}_l(c_0)$ depends on $c_0$ through $\alpha_{S,l}$ and $\alpha_{\textit{diff,l}}$, and allows us to calculate the expected value at the assumed concentration of $\beta$-lactoglobuline of our solution. The physicochemical parameters of $\beta$-lactoglobulin used to apply this model to the present case are provided by \cite{Wahlgren1997SimpleB} and \cite{Rabe2007A-lactoglobulin} and reported in Table \ref{tab:surf}.

\begin{figure}
    \centering
    \includegraphics[width=0.6\linewidth]{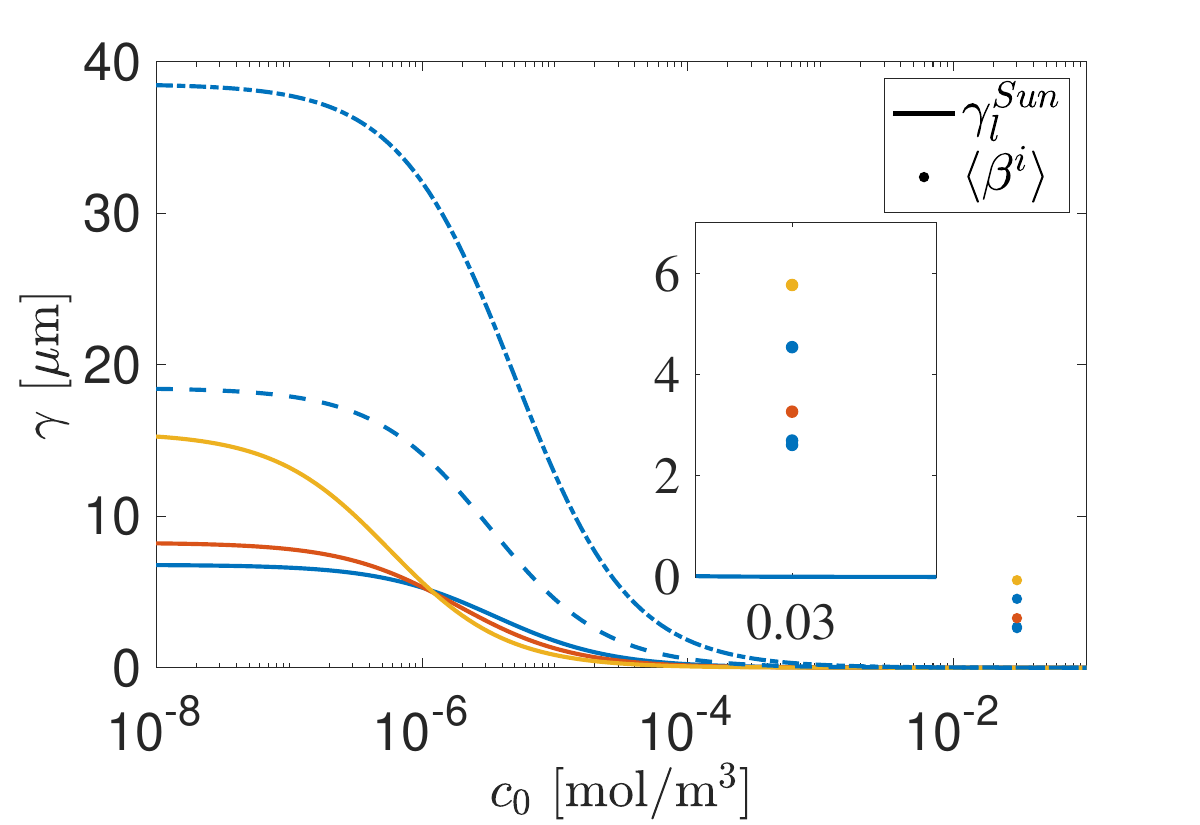}
    \caption{Effective slip lengths $\gamma^{\textit{Sun}}_l$ predicted by Sundin's model as a function of the concentration $c_0$ in $\beta$-lactoglobuline (lines) and spanwise-averaged slip lengths measured at the interface $\langle\beta^i\rangle$ at the present estimated concentration (markers). The colours associated with the cases are: blue for A106N03, A106N1, A106N13; orange for A089N03 and yellow for A059N03. Solid lines represent cases at $N=0.3$, the dashed line represents $N=1.2$ and the dash-dotted line $N=13.2$.}
    \label{fig:betaeff_tauma}
\end{figure}

Figure \ref{fig:betaeff_tauma} shows $\gamma^{\textit{Sun}}_l$ as a function of the bulk concentration of $\beta$-lactoglobulin (lines) compared to the spanwise averaged slip length measured at the interface (points). For all our geometries, the model predicts $\gamma^{\textit{Sun}}_l\approx 0$ for bulk concentrations above $c_0 \sim \SI{1e-5}{\mol\per\meter\cubed}$. This threshold value is considerably lower than the estimated concentration of $\beta$-lactoglobulin for the working fluid, $c_{0,w}=\SI{0.03}{\mol\per\meter\cubed}$. The model thus indicates a total immobilization of the interface. In other words, the model predicts that the Marangoni stress at the interface completely dominates over the lubricant stress, i.e.~$\tau_w = \tau_l + \tau_{Ma} \simeq \tau_{Ma}$. 

Note however that we measure non-zero slip lengths (Figure \ref{fig:localSL_SV}b), which means that the interface retain 
more mobility than predicted by Sundin's model adapted to longitudinal grooves.
It is not clear how the flow of surfactants at the interface behaves in this system and how it relates to adsorption and desorption dynamics. In fact, the experimental surfactant flow field (including surfactants at interfaces and in the bulk) can be more complex than what is assumed in the simplified geometry of the model \citep{Baier2023InfluenceGrooves}. 
A deeper understanding would require further investigations under controlled chemical conditions as well as accurate observation of the stagnation points. 
The model by \cite{Sundin2022SlipSurfactants} nevertheless indicates that the reduced slip length, relative to the idealised LIS model, may result from surfactants in the system. To test this hypothesis, numerical simulations are performed, applying boundary conditions for either a clean mobile interface or a surfactant-laden immobile interface to the experimental configurations. 

\begin{table}
  \begin{center}
\def~{\hphantom{0}}
  \begin{tabular}{llc}
    Variable & Definition & Value \\[3pt]
    Bulk Péclet number & $\textit{Pe}_L=\dfrac{UL}{D_s}$ & $1.2\times 10^{7}$\\ [8pt]
    Interfacial Péclet number & $\textit{Pe}'_s=\dfrac{wu_{\textit{s,SHS}}^0\beta_{\textit{LIS}}}{D_s} $ & 890\\ [8pt]
    Marangoni number & $\textit{Ma}=\dfrac{nRT}{\omega\mu_w U}$ & 254 \\ [8pt]
    Biot number & $\textit{Bi}'_L = \dfrac{\kappa_d L^2}{wu^0_{\textit{s,SHS}}\beta_{\textit{LIS}}}$ & 5.5\\ [8pt]
    Damköhler number & $\textit{Da}_{\delta,L}=\dfrac{k_a\delta_L}{\omega D_s}$ & 0.69\\ [8pt]
    Interfacial diffusivity number & $\alpha_{\textit{diff},l} =\frac{1}{\Pen'_{S} \textit{Ma} c_k}$ & $8.3 \times 10^{-10}$\\ [8pt]
    Surfactant solubility number & $\alpha_{S,l}=\frac{1}{8} \frac{L}{w} \frac{\textit{Bi}'_L}{1+\textit{Da}_{\delta,L}} \frac{1}{\textit{Ma} c_k}$ & $1.0 \times 10^{-4}$\\ [8pt]
  \end{tabular}
  \caption{Non-dimensional numbers used for the definition of the effective slip length $\gamma_l^{\textit{Sun}}$ for a longitudinal LIS in the presence of surfactants. The reported values are calculated for case A106N03. These numbers are in the same order of magnitude for the other four cases.}
  \label{tab:non-dim-numb}
  \end{center}
\end{table}
\begin{table}
  \begin{center}
\def~{\hphantom{0}}
  \begin{tabular}{lll}
   Variable & Notation & Value \\ [3pt]
   Bulk/surface diffusion constant & $D_s$ $[\si{\meter\squared\per\second}]$ & $1.8\times10^{-11}$ \\
   Surfactant concentration in water & $c_0$ $[\si{\mole\per\meter\cubed}]$ & 0.03 \\
   Surfactant adsorption constant & $k_a$ $[\si{\meter\cubed\per\mole\per\second}]$ & 2.3 \\
   Surfactant diffusion constant & $k_d$ $[\si{\per\second}]$ & $1.4\times 10^{-5}$  \\
   Ionic constant & $n$ & 2\\
   Universal gas constant & $R$ $[\si{\joule\per\mole}]$ & 8.314 \\
   Surfactant concentration at interface & $\Gamma_m$ $[\si{\mole\per\meter\squared}]$ & $1.4\times 10^{-7}$ \\
  \end{tabular}
  \caption{Physico-chemical parameters of $\beta$-lactoglobulin used for the calculation of $\gamma_l^{\textit{Sun}}$ (from \protect\cite{Wahlgren1997SimpleB} and \protect\cite{Rabe2007A-lactoglobulin}).
  }
  \label{tab:surf}
  \end{center}
\end{table}

\section{Numerical simulations}
\label{sec:numerics}

In this section, the numerical setup and methodology are presented for modelling interface mobility. The obtained results are combined with the experimental measurements to better understand the effects of surfactants.

\subsection{Configuration and method}

\begin{figure}
    \centering
    \includegraphics[width=0.3\linewidth]{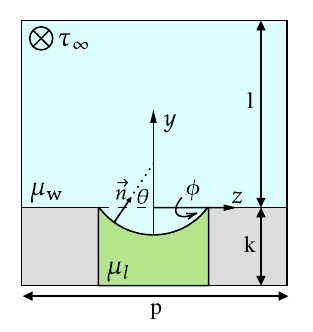}
    \caption{Schematic representation of the numerical setup.}
    \label{fig:DNS_setup}
\end{figure}

We solve the Stokes equations for the two-phase configuration shown in Figure \ref{fig:DNS_setup}. This domain represents a unit cell of the experimental configuration shown in Figures \ref{fig:setup} and \ref{fig:cross_sect}. 
The 2D domain has a width equal to the pitch $p$, $k$ is defined as in Figure \ref{fig:setup} and $l = p - k$ is the distance between the groove crest and the top boundary. Moreover, $\theta$ is the angle that the local normal to the interface $\Vec{n}$ forms with the vertical axis and $\phi$ is the deflection angle of the meniscus (also shown in Figure \ref{fig:cross_sect}). 
Periodic boundary conditions are applied in the spanwise direction, effectively simulating an infinite array of grooves. Moreover, no-slip boundary conditions are imposed on the solid surface. To drive the flow, a constant shear $\tau_\infty$ is applied at the top boundary. 
The variables are non-dimensionalized using the surface tension $\sigma$, fluid density $\rho$, and characteristic length scale $l$, matching the order of magnitude of the dimensionless groups from the experiments: $\Rey = \rho U l/\mu_w$, $\textit{We} = \rho U^2 l/\sigma$, and $\textit{Ca} = U \mu_w/\sigma$, where $U$ is the velocity at $y = l$. Geometric properties ($a$, $A$) and the viscosity ratio $N$ are also matched with the experiments.  The two fluids, water and lubricant, have the same densities.
Since the Capillary and Weber numbers are small ($\textit{Ca} \approx 10^{-4}$ and $\textit{We} \approx 10^{-4}$), the interface is essentially static under the flow. It is initialized with a shape extracted from the OCT measurements (i.e. Figure \ref{fig:cross_sect}).
%

Numerical simulations were conducted using Basilisk \citep{Popinet2003Gerris:Geometries,Popinet2009AnFlows,Popinet2015AEquations}, an open-source framework for solving partial differential equations on adaptive Cartesian meshes.
The interface was tracked through a geometric volume-of-fluid (VOF) method \citep{Scardovelli1999DIRECTFLOW}. 
Despite the low Reynolds number, we solve the time-dependent equations, as it provides a more stable algorithm.
The equations are discretized using a time-staggered approximate projection method on a Cartesian grid. The viscous diffusion term is treated via a second-order Crank–Nicholson fully implicit scheme, while spatial discretization employs second-order finite volume on octree grid. The solid wall is modelled using an immersed boundary (cut-cell) method \citep{Johansen1998ADomains,Mittal2005IMMERSEDMETHODS,Schwartz2006ADimensions}.
The computational domain is a cubic volume with a side length equal to the pitch $p$ of the investigated configuration, with periodic boundary conditions in the spanwise and streamwise directions, no-slip condition at the bottom boundary and the solid wall, and Neumann condition at the top one. A 2D slice of the domain is shown in Figure \ref{fig:DNS_setup} and represents the physical domain under consideration. 
The mesh refinement is based on a tree-grid discretization, with cell dimensions calculated as the domain length $p$ divided by $2^{\text{n}_{\text{ref}}}$, where $\text{n}_{\text{ref}}$ denotes the refinement level. The mesh refinement ranges from level 5 in the far field to level 8 near the interface and solid wall, ensuring that the cell size is at least ten times smaller than the maximum local slip length. This refinement strategy is validated through a mesh independence study (see Appendix \ref{appD}).

The numerical method is validated by simulating the idealized case of a longitudinal groove with a flat interface, as considered in the models of \cite{Schonecker2013LongitudinalFluid} and \cite{Schonecker2014InfluenceState}. The effective slip lengths obtained numerically are compared to the analytical results and reported in Appendix \ref{appD}.

\subsection{Comparison between experiments and simulations}\label{sec:dns_exp}
To test the hypothesis of partial immobilisation of the liquid-liquid interface, the experimental configurations are reproduced numerically, providing the full velocity fields above and within the cavity.
The boundary conditions at the liquid-liquid interface in the simulations are tuned to represent two scenarios: (i) immobilized surfactant-laden interfaces and (ii) surfactant-free interfaces.
In scenario (i), the immobilized regime is modelled by imposing zero velocity at the curved liquid-liquid interface. In scenario (ii), the continuity of velocity and tangential stress at the interface is enforced implicitly through the VOF to allow for local slip.

\begin{figure}
  \centerline{\includegraphics[width=1\textwidth]{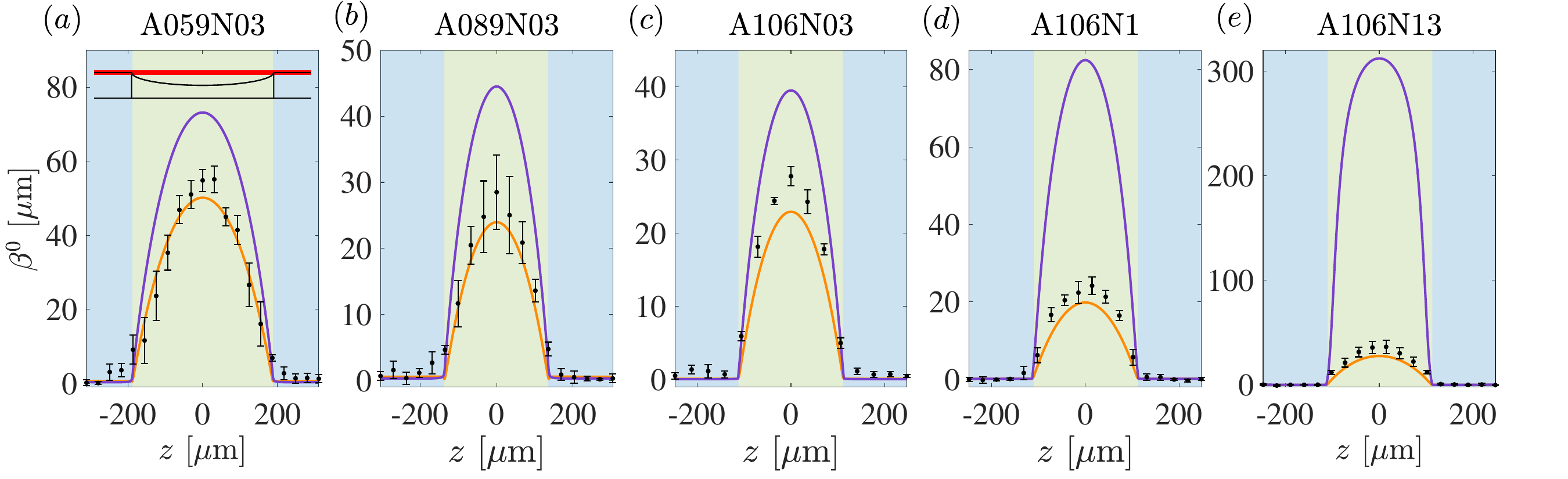}}
  \caption{Experimental slip lengths $\Expz$ and the DNS results at the reference plane ($x,y=0$) with no slip (orange line) and slip (purple line) boundary condition at the interface. The sketch in frame ($a$) represents the reference plane.}
\label{fig:DNS_noslip}
\end{figure}

From the velocity fields, the local slip length distribution is computed by taking the ratio between the local normal shear rate and the velocity along a reference plane  (as detailed in Appendix \ref{appF}).  Since case (i) has, by definition, zero velocity at the liquid-liquid interface, we choose the reference plane to coincide with the crest plane $y=0$ (shown in Figure \ref{fig:DNS_noslip}(a)).   
The slip lengths obtained from a simulation in scenarios (i) and (ii) are respectively denoted by $\DNSns$ and $\DNSs$, where the subscripts $NS$ and $S$ refer to "no-slip" and "slip".
The local slip length distributions from the simulations are compared to the experimental slip length obtained from Equation \eqref{eq:beta_loc_exp}. Note that experimental slip lengths are also evaluated at reference plane $y=0$. The experimental slip length evaluated there is denoted by $\Expz$.

Figure \ref{fig:DNS_noslip} compares the slip lengths extracted from the simulations with those from the experiments. 
The five frames correspond to the five experimental configurations listed in Table \ref{tab:LIS_cases}. The experimental results are represented with black symbols, while the simulated results are shown with solid lines (orange for case (i) and purple for case (ii)). 
For all cases, the experimental values $\Expz$ lie between the two predictions, but much closer to the immobilized case (i).
The difference between maximum values of $\Expz$ and $\DNSns$ is smaller than \SI{20}{\percent} for all cases, which is remarkable, given the simplifications inherent in the numerical model. In contrast, the slip lengths simulated with a fully mobile interface, $\DNSms$ are much larger than the experimental values. 
Furthermore, the difference between the numerical distributions with slip and no-slip conditions is significantly larger at higher viscosity ratios (cases A106N1 and A106N13). This is expected, given that the slip length in the absence of surfactants increases with $N$ (see Figure \ref{fig:valid_N_long} of Appendix \ref{appD}).

To gain further insight into how geometrical features affect the slip length, the average slip length is calculated and related to the aspect ratio of the grooves. The average slip length is defined as the arithmetic mean of the local slip length over a pitch (see Appendices \ref{appE} and \ref{appF}), i.e. 
\begin{equation}
    \langle \beta \rangle = \frac{1}{n}\sum_n\beta(z_n) \qquad \text{with} \qquad -\frac{p}{2} < |z_n| < \frac{p}{2}
\end{equation}
Table \ref{tab:mean_beta} reports the average slip lengths of the LIS computed over the pitch (that is taking into account both the liquid-liquid and liquid-solid interfaces) and we observe again a good match between the experimental values $\Expm$ and $\DNSmns$.

At fixed groove geometry, increasing the viscosity ratio (more viscous working fluid and/or less viscous lubricant) tends to get the system closer to the immobilized case (i). Increasing $N$ by nearly a factor 50 ($ N$ goes from 0.3 to 13) results in a \SI{30}{\percent} increase of the average slip length in the presence of surfactants, whereas in the absence of contaminants, the increase is \SI{88}{\percent}. 

\subsection{Discussion}
For case (i), which models a fully immobilized interface, Marangoni stress dominates $\tau_{\textit{Ma}}\approx\tau_w$, whereas for the surfactant-free case (ii) $\tau_{\textit{Ma}}=0$. These two limiting scenarios can be used to estimate the Marangoni stresses present in the experiments using the model by \cite{Sundin2022SlipSurfactants}. In their model, the effective slip is linearly dependent on the Marangoni stress. Assuming that this is also valid for the average slip length, we may fit a linear function using the two limiting values and estimate the stress ratio $ \tau_{Ma}/\tau_w$ given measurements of the average slip length. 

The last column of Table \ref{tab:mean_beta} reports the stress ratio for the five experimental cases. It can be observed that, the stress ratio $\tau_{\textit{Ma}}/\tau_w$ ranges between $0.7$ to unity.
From the tangential stress balance at the interface, 
\begin{equation}
    \frac{\tau_{\textit{Ma}}}{\tau_w}=1-\frac{\tau_l}{\tau_w}.
    \label{eq:stress-balance-discussion}
\end{equation}
The estimates confirm that for all configurations, the Marangoni stress is the main source of resistance at the interface, i.e. it dominates over the resistance imposed by the lubricant stress.
The ratio between the stress exerted by the water and that exerted by the lubricant is proportional to the viscosity ratio, i.e. ${\tau_l}/{\tau_w}\propto{\mu_l}/{\mu_w}={1}/{N}$. Therefore from \eqref{eq:stress-balance-discussion}, we expect a larger Marangoni stress with increasing $N$, which is also confirmed in Table~\ref{tab:mean_beta}. Note that in section \ref{sec:surfactant} we estimated the Marangoni stress by estimating the lubricant shear stress as $\tau_l \approx 4 \mu_l u_s/k$. The estimates using this approach (last column in Table~\ref{tab:beta_comparison}) are in good agreement with the estimate made here through a calibration with limiting values of average slip length obtained from numerical simulations (last column in Table~\ref{tab:mean_beta}).

It is important to discuss the measured average slip length $\langle \beta^0\rangle $ in Table \ref{tab:mean_beta}, which, despite the minimum mobility of the liquid-liquid interface, amounts to 10 to \SI{20}{\micro\meter}. The reported values represent the average slip at the crest plane of the ridges, i.e. the slip velocity is evaluated in the water phase rather than at the interface of the two fluids. This expedient was previously used by \cite{Crowdy2017EffectiveSurfaces} who treated a similar problem as ours mathematical. Crowdy studied a laminar flow over longitudinal grooves where the interface is immobile and the interface has a meniscus deflected by an angle $\phi$ from the plane $y=0$. To derive the explicit expression of the effective slip length, Crowdy used a conformal mapping method to map the curved interface into the plane of the ridges. The simulated slip lengths of a completely immobile interface are in good agreement with Crowdy's predictions (see Figure \ref{fig:crowdy}$a$ in Appendix). Another analytical approach accounting for the presence of surfactant is the model of \cite{Baier2023InfluenceGrooves}, where the curved surfactant-laden interfaces are not immobilized but rather incompressible. In this case, the velocity profile at the interface is non-zero with a zero average across the groove, thus predicting the presence of interfacial back-flows. This model explores other ways to add surfactants as a physical ingredient in LIS in a more refined way than immobilizing the interface. The average slip-length prediction remains the same as in the immobilized case of \cite{Crowdy2017EffectiveSurfaces} (that is due to the presence of the meniscus). The predicted recirculations are not observed in our experiments. 

\begin{table}
  \begin{center}
\def~{\hphantom{0}}
  \begin{tabular}{lcccc}
    Case & $\Expm$ [\unit{\micro\meter}] & $\DNSmns$ [\unit{\micro\meter}]& $\DNSms$ [\unit{\micro\meter}] & $\tau_{\textit{Ma}}/\tau_w$\\ [3pt]
    A059N03 & 21.42 $\pm$ 1.33 & 20.82 & 31.19 & 0.91 \\
    A089N03 & 8.70 $\pm$ 0.19 & 7.49 & 14.17 & 0.84 \\
    A106N03 & 8.57 $\pm$ 1.63 & 7.04 & 12.44 & 0.69 \\
    A106N1 & 7.56 $\pm$ 0.71 & 6.09 & 26.22 & 0.93 \\
    A106N13 & 11.34 $\pm$ 1.89 & 8.51 & 106.36 & 0.97 \\
  \end{tabular}
  \caption{Average slip lengths derived from experiments $\Expm$ compared to the numerical results in case of immobile $\DNSmns$ and slipping $\DNSms$ interface.}
  \label{tab:mean_beta}
  \end{center}
\end{table}

\section{Conclusions}
\label{sec:conclusion}
For the first time, the local slip length over LISs with longitudinal grooves exposed to laminar flow has been measured through direct local velocity profiling. Using Doppler-optical coherence tomography, the shape of the liquid-liquid interface and the local velocity were measured with a micrometre spatial resolution. To investigate the role of the groove width and viscosity ratio on slip, five LISs configurations were constructed.
The comparison between the measured slip lengths and the predictions of established analytical models \citep{Schonecker2013LongitudinalFluid}, provided the starting point to quantify the influence of curved liquid-liquid interfaces and the presence of surfactants dispersed in the bulk flow.

The experimental cases were compared with numerical simulations where the boundary conditions were tuned to create two limiting scenarios, one of a surfactant-saturated interface and one of a surfactant-free interface. The first scenario corresponds to a rigid (no-slip velocity) corrugated surface, while the second scenario is a surface with finite local slip. Local slip length distributions over a pitch were calculated and compared to the corresponding experimental values.
We measured nearly the same slip lengths experimentally as obtained from simulations of a rigid curved interface, indicating that we have strong resistance from Marangoni stresses, which immobilizes the interface. As explained by \cite{Sundin2022SlipSurfactants}, this stress is due to a gradient in surfactant concentration. The concentration is higher downstream than upstream in the groove resulting in a Marangoni stress that increases the resistance of the interface and thus reduces the slip velocity.
The evidence of contaminants causing this detrimental effect is reinforced by the observation that, for a given groove geometry, the average slip length remains nearly constant and independent of the viscosity ratio. This suggests that the interface becomes more rigid, preventing the lubricating fluid from influencing the upper fluid as effectively. Surface-active contaminants can therefore be responsible for a failure mechanism of LISs in applications where enhanced interfacial slip is required, such as drag reduction. 

Our study indicates that a careful design of a LIS for a given working fluid and application is necessary to limit the effect of contaminants. Further studies are required to understand how surface texture and lubricant can be tuned to counteract, or at least minimize, the Marangoni stress. The presence of surfactant is not the main issue, it is the concentration gradient that needs to be counteracted, which may be achieved through the design of geometry under given flow conditions. 
%
For example, the formation of surfactant concentration gradients can be modified by changing the stagnation points (e.g. barriers in the groove that stagnate the lubricant flow) of a LIS geometry. 

An additional contribution of our study is the first use of D-OCT as a relatively accurate technique to directly measure the slip velocity on a LIS and monitor the lubricant coverage at the same time.
However, the requirement of an opaque fluid to perform the measurements limits the range of materials this method can be used with.  
For instance, D-OCT is well-suited for studying slip in blood flows or polymer suspensions, where the fluid’s optical density eliminates the need for additional scattering agents. Future work will use tracer particles that are not surface active, to determine the effect of naturally occurring surfactants, as well as, to more quantitively determine their influence on slippage.

\backsection[Acknowledgements]{We are grateful to Zhuxuan Cui for the fabrication of the silicon moulds.}

\backsection[Funding]{
The authors acknowledge financial support from the
European Union through the European Research Council
(ERC) grant no. “CoG-101088639 LUBFLOW”, Knut and Alice Wallenberg Foundation (KAW) grant no. 2016.0255 and the Swedish Foundation for Strategic Research (SSF) grant no. FFL15:0001. Access to the computational resources used for this work was provided by the Swedish National Infrastructure for Computing (NAISS).
}

\backsection[Declaration of interests]{The authors report no conflict of interest.}


\appendix

\section{LIS assembly and groove fabrication}\label{appA}
\subsection{Assembly of flow cell}
\label{app:A1}
\begin{figure}
    \centering
    \includegraphics[width=0.6\linewidth]{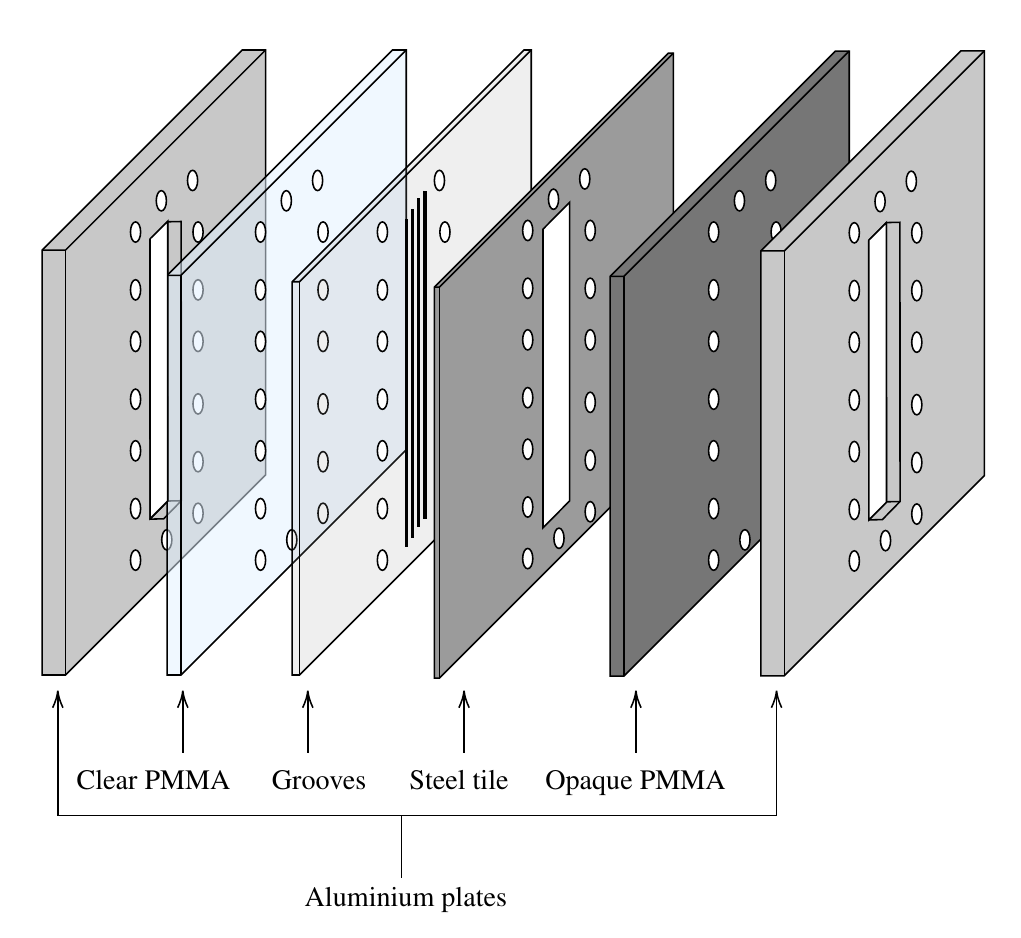}
    \caption{Schematic of the assembly of the flow cell.}
    \label{fig:lis_assembly}
\end{figure}

As illustrated in Figure \ref{fig:lis_assembly}, the flow cell consists of four square panels sandwiched between two aluminium plates, which are fastened by screws to ensure uniform pressure across the entire structure. The panels are arranged from top to bottom of the channel (left to right in Figure \ref{fig:lis_assembly}) as follows: a transparent tile of PMMA, a clear plastic sheet where the ridges are cast, a stainless steel tile with a rectangular slit of dimensions $3.25\times\SI{80}{\milli\meter\squared}$ and \SI{1}{\milli\meter} thick forms the duct, and a smooth hydrophilic tile made of opaque PMMA as the bottom of the duct. Mounting the grooves on the top wall of the duct is crucial for the accurate measurement of the velocity at the liquid-liquid interface. In this way, any spurious velocity measurement that can arise from the reflection of the OCT beam at the bottom wall is avoided. 

\subsection{Surface fabrication}
\label{app:A2}
The substrates of the LISs are fabricated from Ostemer 322 (Mercene Labs, Stockholm, Sweden), Off-Stoichiometry-Thiol-Ene resin (OSTE) via positive soft moulding with a silicon master made by deep ion etching. Components A and B of the resin are mixed with a speedmixer in ratio 1:1.09 as prescribed by the manufacturer and the mixture is poured on the mould. A transparent release liner applied on top facilitates the delamination from the mould and acts as a support layer to the surface. The sample is then exposed to an uncoherent UV-light source for 60 seconds and the solidified resin is removed from the mould. Finally, the sample is heat-cured at $100^\circ$C for one hour. Subsequently, a hydrophobic spray coating (HYDROBEAD-T) is applied to the grooved surface, to ensure that when the surface is immersed, the lubricant wets the surface better than water. In this way, the established design requirements for LISs are satisfied \citep{Wong2011BioinspiredOmniphobicity}.

In each experiment, the LIS was prepared through complete infusion of the duct with lubricant, followed by injection of the working fluid into the flow cell using a syringe pump (AL-1000, WPI) at a volumetric flow rate of $\SI{20}{\micro\liter\per\minute}$ until filled. The preparation procedure was continuously monitored via cross-sectional scanning of the duct with OCT. 
With this procedure, the lubricant is confined within the grooves for the full duct length.

With the chosen geometry and flow conditions, the lubricant layer will not experience shear-driven drainage. This is supported by estimating the maximum length of lubricant retained in the groove under shear flow, $L_{\infty}$ \citep{Wexler2015Shear-drivenSurfaces}. The retention length  is determined by the balance between shear force and capillary forces,
\begin{equation}
    L_{\infty}=\frac{c_p}{c_s}\frac{k}{r_{\textit{min}}}\frac{\sigma}{\tau_w}.
\end{equation}
Here, ${c_p}/{c_s}\approx 1$ represents the ratio of non-dimensional geometric factors, $r_{\textit{min}}$ is the minimum radius of curvature of the interface, $\sigma$ is the interfacial tension and $\tau_w$ is the shear stress (from the overlying flow). In our case, $r_{\textit{min}}\approx\SI{400}{\micro\meter}$, $\sigma\approx\SI{50}{\milli\newton\per\meter}$ and $\tau_w\approx\SI{4}{\milli\pascal}$, resulting in an estimated retention length of $L_{\infty}\approx\SI{4}{\meter}$, far exceeding the length of the flow cell.


\section{Uncertainty analysis}\label{appB}
The uncertainty of the single slip length is calculated by propagating the errors associated with the velocity measurements. Considering equation \eqref{eq:beta_loc_exp}, the overall uncertainty is estimated as follows:
\begin{equation}
    \sigma_{\beta}^2=\left(\frac{\partial\beta}{\partial u_s}\right)^2\sigma_{u_s}^2+\left(\frac{\partial\beta}{\partial\Delta}\right)^2\sigma_{\Delta}^2
\end{equation}
where $\Delta=\partial u/\partial y$ is the velocity gradient evaluated at the interface. The uncertainty in the experimental measurement of the slip velocity, $\sigma_{u_s}$, is calculated as the standard deviation of the 1000 repeated scans needed to perform a single velocity acquisition and amounts on average to \SI{5}{\percent} of the measured value. The velocity gradient $\Delta$, corresponds to the fitting parameter that linearly interpolates the acquired velocity data close to the interface. Consequently, its standard deviation corresponds to the uncertainty derived from the fitting operation, which is \SI{15}{\percent} on average.

In each configuration, the measurement is repeated in five spanwise positions where the interface curvature was rather constant and then their averaged is computed. The error bars shown in Figure \ref{fig:localSL_SV}$b$ correspond to the standard deviations of these repeated measurements, indicating the range in which \SI{68}{\percent} of all the measurements is contained. Since the standard deviation is consistently larger than the propagated errors, this measure is considered as the variation of the experimental conditions.


\section{Sundin's model for longitudinal grooves}\label{appC}

In the transverse case, the effective slip length found with the model of \cite{Sundin2022SlipSurfactants} is (Equation 5.16):
\begin{equation}
    \gamma^{\textit{Sun}}_t = b_{\textit{SHS,t}} \ \beta_{\textit{LIS,t}} \left( 1- \frac{1}{1+\alpha_{\textit{S,t}} + \alpha_{\textit{diff,t}}} \right).
\end{equation}
The parameter $b_{\textit{SHS,t}}$ represents the effective slip length for a clean transverse SHS whose grooves have pitch $p$ and width $w$, derived by \citep{JohnPhilip1972IntegralConditions}. The expression is given by
\begin{equation}
    b_{\textit{SHS,t}}=-\frac{p}{2\pi}\ln{(\cos\alpha)} \quad \text{and} \quad \alpha = \frac{\pi}{2} \frac{w}{p}.
\end{equation}
Moreover, $\beta_{\textit{LIS,t}}=C_t/(1+C_t)$ is the factor introduced by \cite{Schonecker2014InfluenceState} to take into account non-zero viscosity fluids in the grooves in the transverse configuration. Here,
\begin{equation}
    C_t = \dfrac{8\alpha D_t N}{\ln\left(\frac{1+\sin(\alpha)}{1-\sin(\alpha)}\right)}
\end{equation}
with $D_t$ depending only on the geometrical features of the grooves.

The parameters $\alpha_{\textit{diff,t}}$ and $\alpha_{\textit{S,t}}$ come from Equations (5.18-19) of \cite{Sundin2022SlipSurfactants} and are given by,
\begin{equation}
    \alpha_{\textit{diff,t}} = \frac{1}{\Pen'_s \textit{Ma} c_k} \quad \text{and} \quad
    \alpha_{\textit{S,t}} = \frac{1}{8}\frac{\textit{Bi}'}{1+\textit{Da}_\delta}\frac{1}{\textit{Ma} c_k}.
\end{equation}
To define the non-dimensional numbers, the following variables are used in the model: $u_{\textit{s,SHS}}^0$ is the slip velocity at the interface for a superhydrophobic surface, $D_s$ is the bulk/surface diffusion constant, $k_d$ and $k_a$ are respectively the surfactant diffusion and adsorption constants, $\delta_L$ is the width of the surfactant adsorption boundary layer, $U$ is the bulk flow velocity, $\omega=1/\Gamma_m$ is the surfactant coverage fraction, $\Gamma_m$ is the maximum concentration of surfactants at the interface, $n$ is a constant that accounts for the adsorption of counter-ions \citep{chang1995adsorption,Fainerman2019}. The ionic charge of $\beta$-lactoglobulin is in the range from 1 to 13 \citep{menon1998measurement}. In the present case it is estimated $n=2$, considering the contribution of many other ionic species present in milk.  

When $\alpha_{\textit{diff,t}} \gg \alpha_{\textit{S,t}}$, the interfacial surfactant population transport is dominated by diffusion, and when $\alpha_{\textit{diff,t}} \ll \alpha_{\textit{S,t}}$, it is dominated by bulk-interface exchange. They depend on the non-dimensional numbers,

\begin{align}
    \Pen'_s = \frac{wu^0_{s,SHS}\beta_{LIS}}{D_s} & \quad  \text{surface Péclet number;} \\
    \textit{Bi}' = \frac{\kappa_d w}{u^0_{\textit{s,SHS}}\beta_{\textit{LIS}}} &\quad \text{Biot number at the interface;} \\
    \textit{Da}_\delta = \frac{\kappa_a \delta}{D_s} & \quad \text{Damköhler number;}  \\
    \textit{Ma} = \frac{nRT}{\omega\mu_wU}  &\quad \text{Marangoni number;}\\
    c_k = \frac{\kappa_a c_0}{\kappa_d} &\quad \text{equilibrium surfactant interfacial concentration}
\end{align}

This model takes into account the bulk and interface transport of surfactants in a transverse flow above a flat lubricant-infused groove. The contributions of the geometrical features, the viscous effects in the groove and the Marangoni stress induced by surfactant gradients at the interface are all modelled, yielding the expression above for the effective slip length, $\gamma^{\textit{Sun}}_t$.
This model predicts a vanishing slip length for very low concentration of surfactants, even with soluble uncontrollable traces in clean experimental working environments with a bulk surfactant concentration $c_0 \sim 10^{-4} \si{\milli\mol\per\liter}$. 

However in the longitudinal case, the symmetry of the problem changes, and we must take into account the typical length $L$ along the $z$ direction, which changes the derivation of the effective slip length. An important scaling argument is that this length $L \gg w$ can be expected to effectively \textit{dilute} the interfacial surfactant population gradient, as $\omega/L \ll \omega/w$, leading to a much lower Marangoni stress at a given surfactant bulk concentration $c_0$.
To understand if this effect is sufficient for this model to predict a non-zero slip given our configuration and chemistry, we need to re-derive it for the longitudinal configuration.

This first step is an additional assumption: we will suppose that the surfactant population in the bulk, $c_0$, and at the interface, $\Gamma(z)$, vary along $z$, but are now invariant along $x$, contrary to the flow field $u(x,y)$, whose symmetry is still $z$-independent.
This yields the same form for the expression of the effective slip length for the longitudinal case:
\begin{equation}
    \gamma^{\textit{Sun}}_l = b_{\textit{SHS,l}} \ \beta_{\textit{LIS,l}} \left( 1- \frac{1}{1+\alpha_{\textit{S,l}} + \alpha_{\textit{diff,l}}} \right),
\end{equation}
where we redefine the effective slip length for a clean longitudinal SHS as
\begin{equation}
    b_{\textit{SHS,l}}=-\frac{p}{\pi}\ln{(\cos\alpha)}
\end{equation}
with $\alpha=(\pi/2)(w/p)$. The parameter $\beta_{\textit{LIS,l}} = C_l/(1+C_l)$ is the factor introduced by \cite{Schonecker2014InfluenceState} for a viscous lubricant in a longitudinal groove, with
\begin{equation}
    C_l = \frac{2\pi a D_l N}{\ln \left(\frac{1+\sin(\alpha)}{1-\sin(\alpha)}\right)},
\end{equation}
and $D_l$ depending only on the geometrical features of the grooves.

The variables $\alpha_{\textit{diff,l}}$ and $\alpha_{\textit{S,l}}$ represent the effects of interfacial surfactant diffusion and bulk exchange on the slip length, respectively. They are defined as
\begin{equation}
    \alpha_{\textit{diff,l}} = \frac{1}{\Pen'_{s} \textit{Ma} c_k} \quad \text{and} \quad
    \alpha_{\textit{S,l}} = \frac{1}{8} \frac{L}{w}\frac{\textit{Bi}'_L}{1+\textit{Da}_{\delta,L}} \frac{1}{\textit{Ma} c_k}.
\end{equation}
Where we redefined the following non-dimensional numbers:

\begin{align}
    \textit{Bi}'_L = \frac{\kappa_d L^2}{wu^0_{\textit{s,SHS}}\beta_{\textit{LIS}}} &\quad \text{Biot number at the interface;} \\
    \textit{Da}_{\delta,L} = \frac{\kappa_a \delta_L}{D_s} & \quad \text{Damköhler number}.
\end{align}
The length $\delta_L$ is reworked to match the same boundary conditions, depending on the new longitudinal bulk Péclet number $\Pen_L = UL/D_s$ (see Section 5.1 of \cite{Sundin2022SlipSurfactants}). It writes
\begin{equation}
    \delta_L = \frac{1}{2} \frac{L}{1+\left(\frac{L\Pen_L}{4w}\right)^{1/3}}.
\end{equation}
This adaptation of the model provides us with a prediction for $\gamma^{Sun}_l$ in the longitudinal case. It can be computed by providing the geometrical features of the channel as well as the chemical parameters of $\beta$-lactoglobulin (\textit{cf.} Table \ref{tab:surf}).

\begin{figure}
\centering
    \tikzset{every picture/.style={line width=0.75pt}} 
    \raisebox{-1\height}{\begin{tikzpicture}[x=0.75pt,y=0.75pt,yscale=-1,xscale=1]
    \node [inner sep=0pt,below right,xshift=0.0\textwidth] 
                {\includegraphics[width=0.46\textwidth]{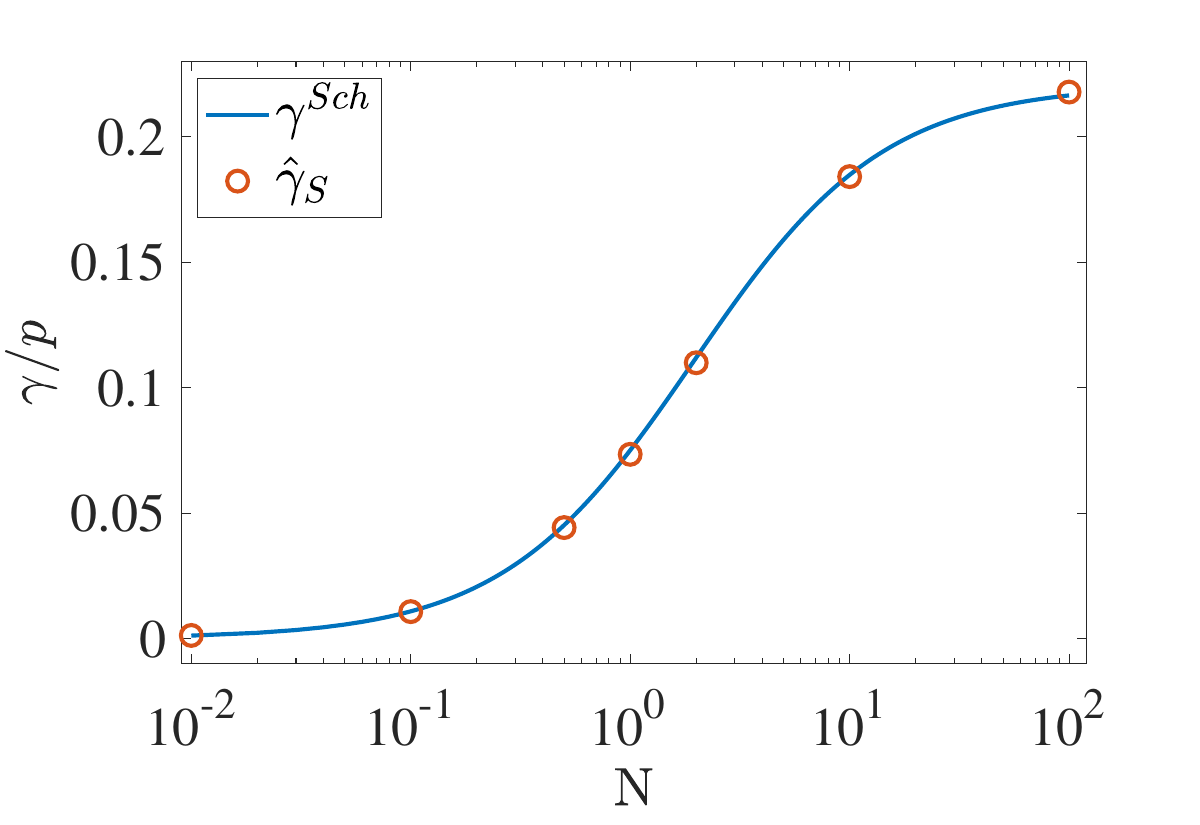}};
    \draw (-10,10) node [anchor=north west][inner sep=0.75pt]   [align=left] {($a$)};
    \end{tikzpicture}}
    \tikzset{every picture/.style={line width=0.75pt}} 
    \raisebox{-1\height}{\begin{tikzpicture}[x=0.75pt,y=0.75pt,yscale=-1,xscale=1]
    \node [inner sep=0pt,below right,xshift=0.0\textwidth] 
                {\includegraphics[width=0.46\textwidth, trim={0 0 0 0},clip]{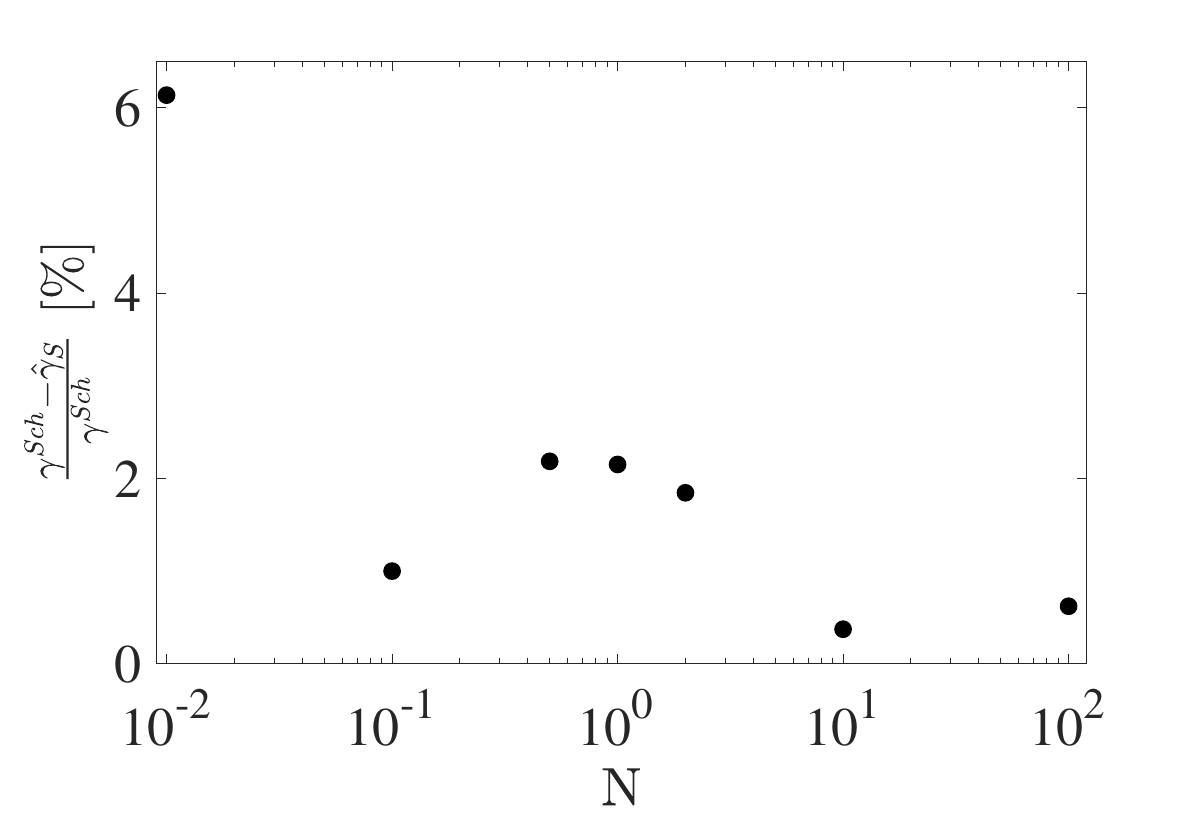}};
    \draw (-10,10) node [anchor=north west][inner sep=0.75pt]   [align=left] {($b$)};
    \end{tikzpicture}} 
    \caption{($a$) Comparison between the normalised effective slip lengths $\gamma/p$ obtained from numerical simulations and the analytical model of \protect\cite{Schonecker2014InfluenceState} as a function of the viscosity ratio $N$. ($b$) Relative error between the two evaluations. In these tests, $A=0.5$ and $a=0.667$.}
    \label{fig:valid_N_long}
\end{figure}

\section{Validation of the numerical setup}\label{appD}
The validation of the numerical simulations was performed by reproducing the configurations considered by \cite{Schonecker2014InfluenceState} and comparing the results with their analytical solutions. 
The validation set up is the same described in section \ref{sec:numerics}, with the angle $\phi = 0$, \textit{i.e.} a flat water-lubricant interface.

In the first validation, a cavity with aspect ratio $A = 0.5$ and slipping area fraction $a = 0.667$ was used, and the viscosity ratio varied from $N = 0.01$ to $N =100$ to assess the quality of the results for seven different cases.
The effective slip length was calculated from the simulated velocity field as 
\begin{equation}
    \hat{\gamma}_S=\frac{Q}{p\frac{\tau_\infty}{\mu_w}H}-\frac{H}{2}
    \label{eq:dns_Q}
\end{equation}
where $Q$ is the volumetric flow rate obtained with $\tau_{\infty}$ applied at the top boundary, and compared with Equation 2.29 in \cite{Schonecker2014InfluenceState}.
Figure \ref{fig:valid_N_long}($a$) shows that the values of the numerical and analytical effective slip lengths normalised with the pitch, $\gamma/p$, are in good agreement. The relative errors between the two evaluations, reported in Figure \ref{fig:valid_N_long}($b$), are below $\SI{3}{\percent}$ for all the tested cases, except at $N=0.01$ where it increases to $\SI{6}{\percent}$.

\begin{figure}
\centering
    \tikzset{every picture/.style={line width=0.75pt}} 
    \raisebox{-1\height}{\begin{tikzpicture}[x=0.75pt,y=0.75pt,yscale=-1,xscale=1]
    \node [inner sep=0pt,below right,xshift=0.0\textwidth] 
                {\includegraphics[width=0.46\textwidth]{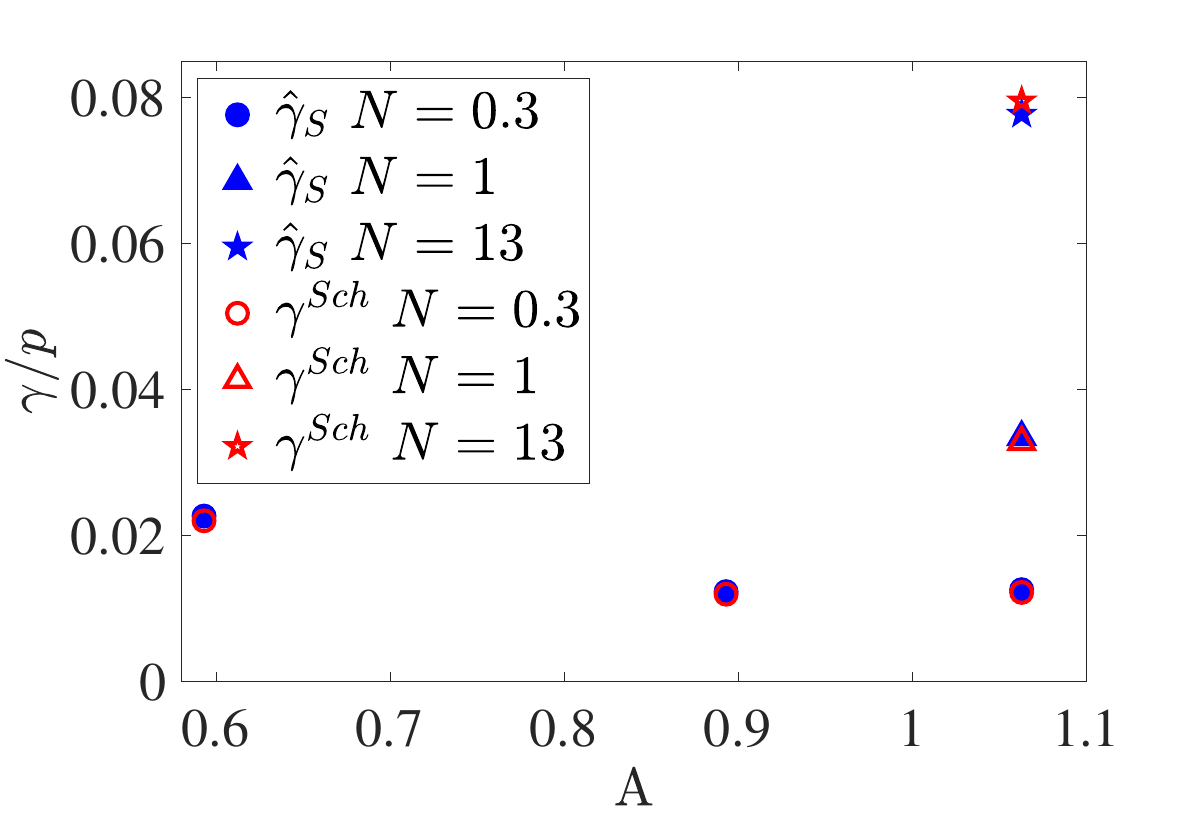}};
    \draw (-10,10) node [anchor=north west][inner sep=0.75pt]   [align=left] {($a$)};
    \end{tikzpicture}}
    \tikzset{every picture/.style={line width=0.75pt}} 
    \raisebox{-1\height}{\begin{tikzpicture}[x=0.75pt,y=0.75pt,yscale=-1,xscale=1]
    \node [inner sep=0pt,below right,xshift=0.0\textwidth] 
                {\includegraphics[width=0.46\textwidth, trim={0 0 0 0},clip]{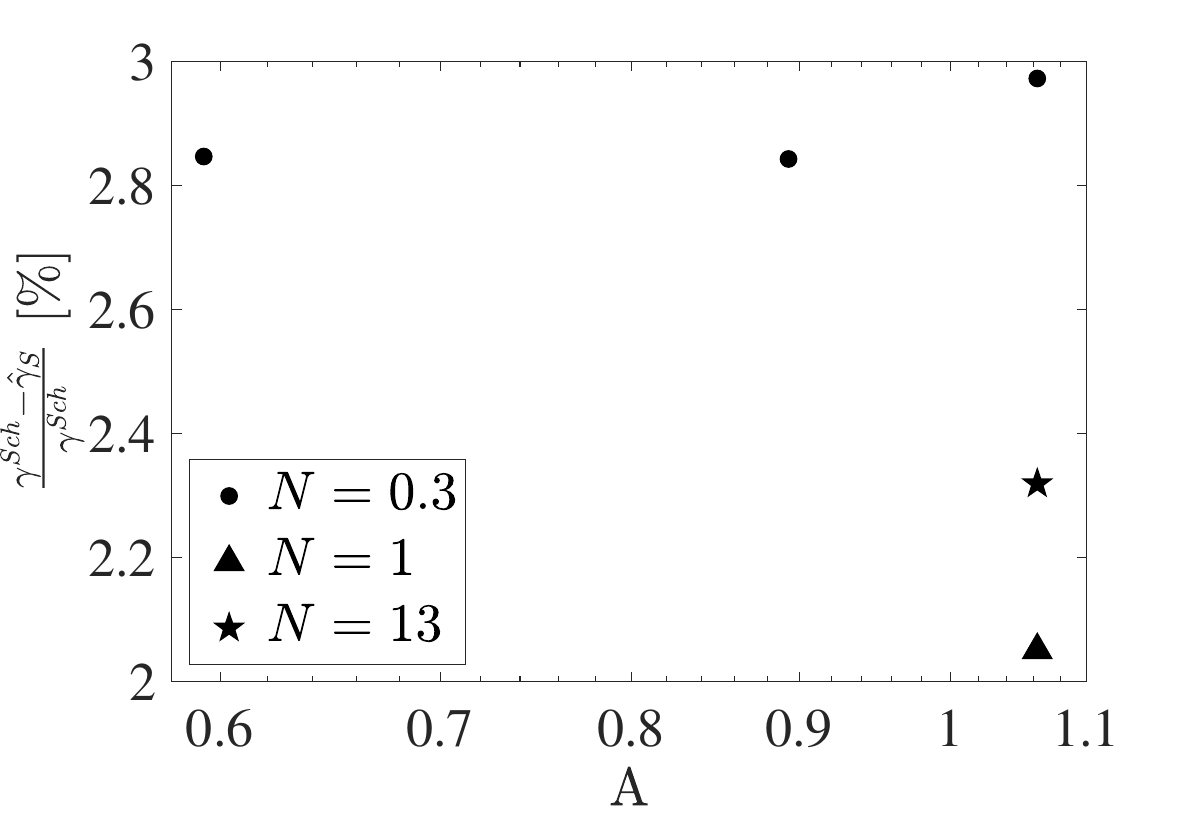}};
    \draw (-10,10) node [anchor=north west][inner sep=0.75pt]   [align=left] {($b$)};
    \end{tikzpicture}} 
    \caption{($a$) Comparison between the normalised effective slip lengths $\gamma/p$ obtained from numerical simulations and the analytical model of \protect\cite{Schonecker2014InfluenceState} as a function of the aspect ratio $A$. ($b$) Relative error between the two evaluations. The geometrical dimensions of the cavity and the viscosity rations reproduce the five experimental cases and the interface is flat.}   
    \label{fig:validation_num}
\end{figure}

Similarly, the validation test is repeated with the dimensions of the cavity aligned with those of the groove dimensions employed in the experimental configurations, as well as the viscosity ratios ($N=0.3, 1.2, 13.2$). Also in this case the interface between the two fluids is flat. We find again good agreement between the simulated effective slip lengths and the analytical predictions, as shown in Figure \ref{fig:validation_num}$a$, with a small relative error (Figure \ref{fig:validation_num}$b$).

\begin{figure}
\centering
    \tikzset{every picture/.style={line width=0.75pt}} 
    \raisebox{-1\height}{\begin{tikzpicture}[x=0.75pt,y=0.75pt,yscale=-1,xscale=1]
    \node [inner sep=0pt,below right,xshift=0.0\textwidth] 
                {\includegraphics[width=0.46\textwidth]{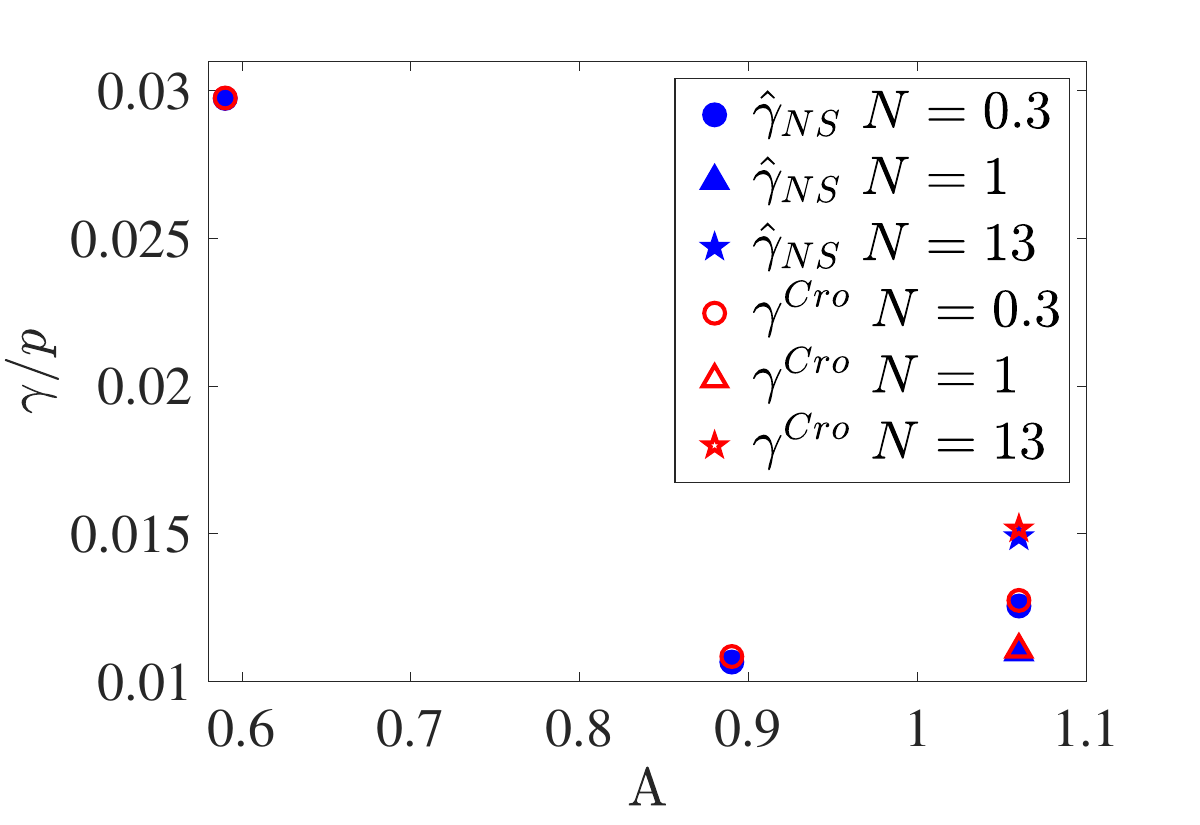}};
    \draw (-10,10) node [anchor=north west][inner sep=0.75pt]   [align=left] {($a$)};
    \end{tikzpicture}}
    \tikzset{every picture/.style={line width=0.75pt}} 
    \raisebox{-1\height}{\begin{tikzpicture}[x=0.75pt,y=0.75pt,yscale=-1,xscale=1]
    \node [inner sep=0pt,below right,xshift=0.0\textwidth] 
                {\includegraphics[width=0.46\textwidth, trim={0 0 0 0},clip]{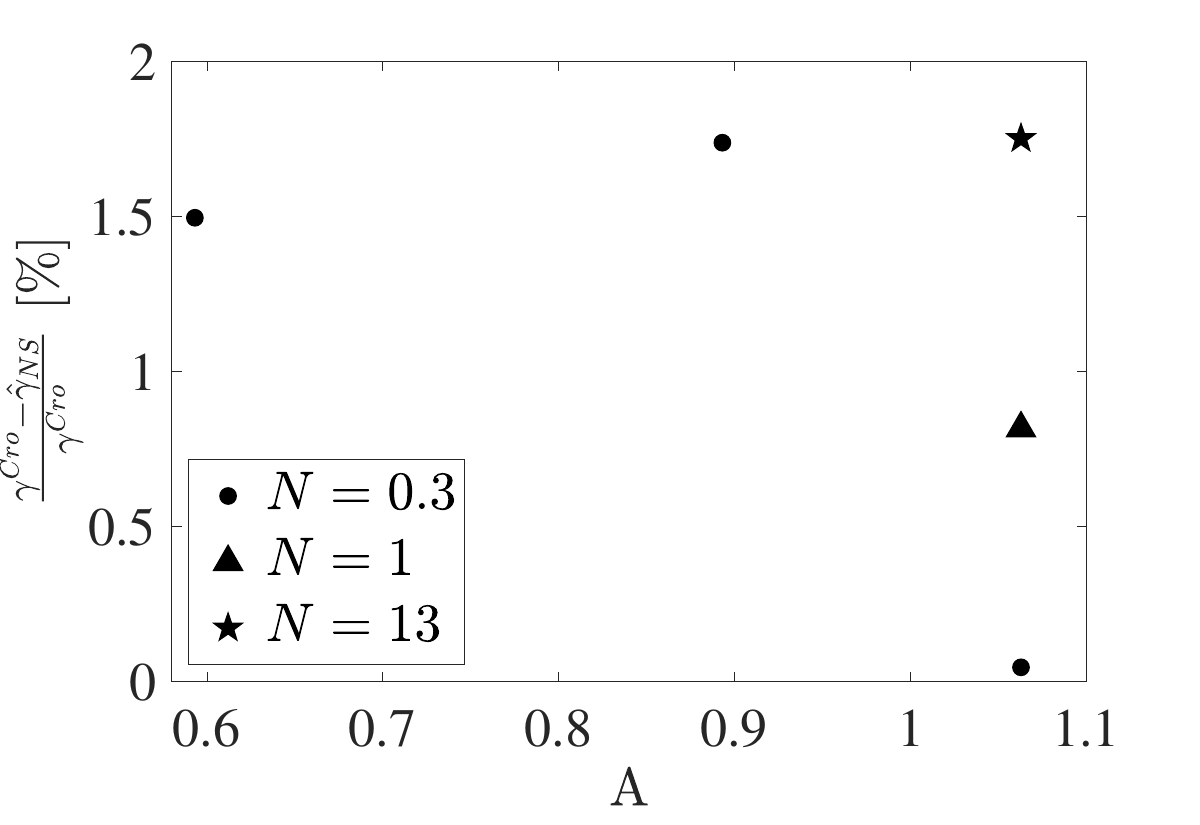}};
    \draw (-10,10) node [anchor=north west][inner sep=0.75pt]   [align=left] {($b$)};
    \end{tikzpicture}} 
    \caption{($a$) Comparison between the normalised effective slip lengths $\gamma/p$ obtained from numerical simulations and the analytical model of \protect\cite{Crowdy2017EffectiveSurfaces} as a function of the aspect ratio $A$. ($b$) Relative error between the two evaluations. The geometrical dimensions of the cavity and the viscosity rations reproduce the five experimental cases with a curved interface.}   
    \label{fig:crowdy}
\end{figure}

The simulated velocity field and the relative effective slip lengths in the scenario of an immobilised interface with a curved meniscus are validated through the comparison of the numerical results with the analytical derivation of \cite{Crowdy2017EffectiveSurfaces}. Equation 1.5 in \cite{Crowdy2017EffectiveSurfaces} was used to calculate the predicted value of the effective slip length, $\gamma^{\textit{Cro}}$, where the geometrical features of the experiments (groove dimensions and meniscus deflection) have been matched. Good agreement between the effective slip lengths obtained from simulations and theory can be observed in Figure \ref{fig:crowdy}$a$ and the small relative errors are shown in Figure \ref{fig:crowdy}$b$.

\begin{figure}
    \centering
    \includegraphics[width=1\linewidth]{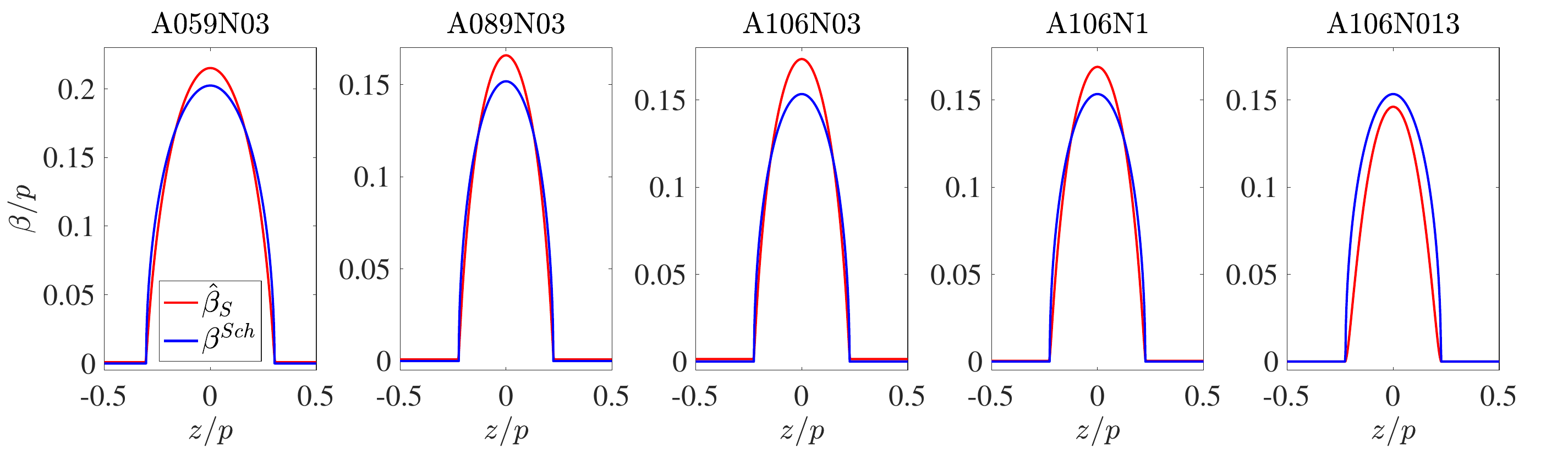}
    \caption{Local slip length comparison between the numerical result and the analytical model}
    \label{fig:local_val}
\end{figure}

The final assessment entails a comparison of the normalised local slip length distributions, $\beta/p$, obtained numerically and analytically, with the same configurations used in the experiments. The local slip length is calculated from the numerical velocity field as 
\begin{equation}
    \frac{\hat\beta}{p}=\frac{u_s}{\frac{\partial u}{\partial y}|_{y=0}}
\end{equation}
where the slip velocity $u_s$ is evaluated at $y=0$.

The analytical expression used to calculate $\beta^{\textit{Sch}}/p$ as comparison is given by Equation 2.28 in \cite{Schonecker2014InfluenceState}. The results are shown in Figure \ref{fig:local_val} where each of the five frames corresponds to a test case. The trend and magnitude of the predicted local slip lengths are reproduced by the numerical model with good agreement.

\begin{figure}
\centering
    \tikzset{every picture/.style={line width=0.75pt}} 
    \raisebox{-1\height}{\begin{tikzpicture}[x=0.75pt,y=0.75pt,yscale=-1,xscale=1]
    \node [inner sep=0pt,below right,xshift=0.0\textwidth] 
                {\includegraphics[width=0.46\textwidth]{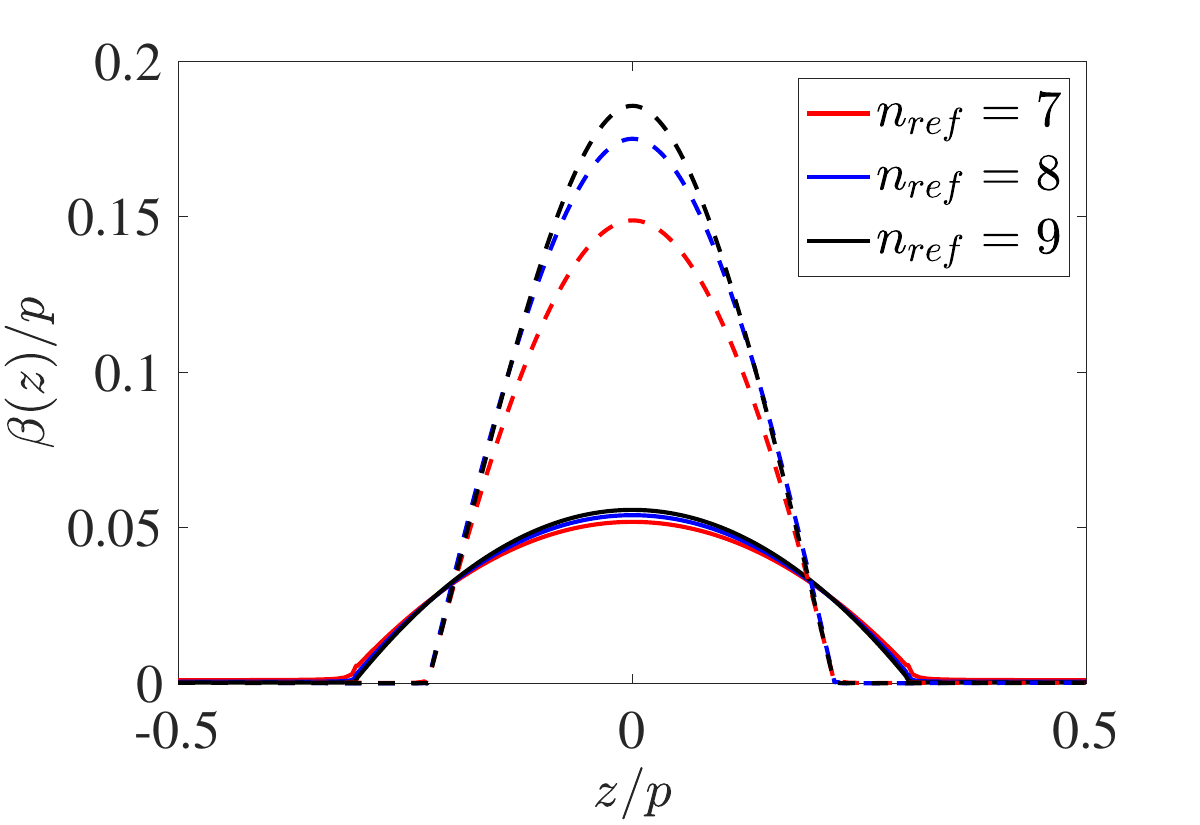}};
    \draw (-10,10) node [anchor=north west][inner sep=0.75pt]   [align=left] {($a$)};
    \end{tikzpicture}}
    \tikzset{every picture/.style={line width=0.75pt}} 
    \raisebox{-1\height}{\begin{tikzpicture}[x=0.75pt,y=0.75pt,yscale=-1,xscale=1]
    \node [inner sep=0pt,below right,xshift=0.0\textwidth] 
                {\includegraphics[width=0.46\textwidth, trim={0 0 0 0},clip]{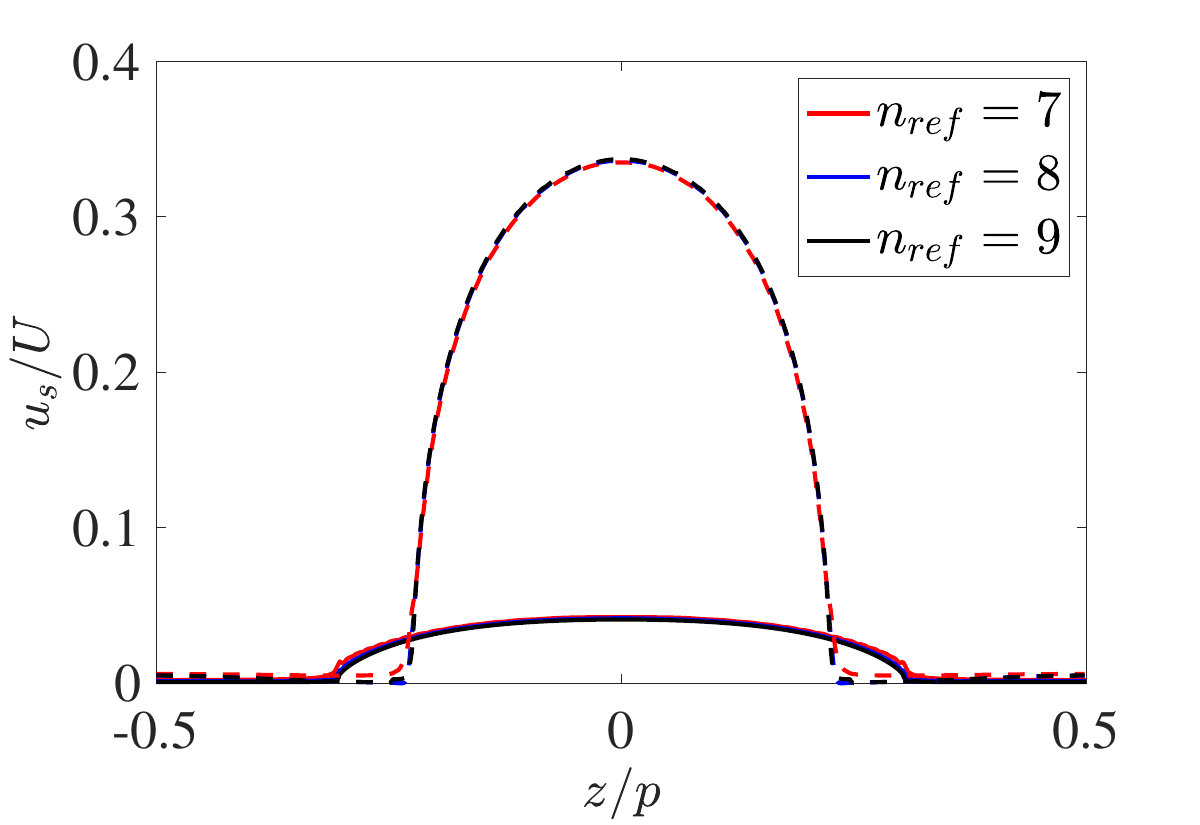}};
    \draw (-10,10) node [anchor=north west][inner sep=0.75pt]   [align=left] {($b$)};
    \end{tikzpicture}} 
    \caption{Mesh convergence study for three different levels of refinement. Two cases are shown, i.e $A=0.59$ with $N=0.3$ (solid lines) and $A=1.06$ with $N=13$ (dashed lines). ($a$) Local slip length. ($b$) Local slip velocity}   
    \label{fig:mesh_independence}
\end{figure}

Regarding the refinement level used for the simulations, a mesh independence analysis was conducted, as illustrated in Figure \ref{fig:mesh_independence}. The figure shows the local slip length ($a$) and slip velocity ($b$) for two distinct cases involving a meniscus, namely $A=0.59$ with $N=0.3$ (solid lines) and $A=1.06$ with $N=13$ (dashed lines) obtained with three different levels of refinement $n_{\textit{ref}}=7,8,9$ that correspond to a minimum cell size of, respectively, $p/2^7$, $p/2^8$ and $p/2^9$. While both the slip length and the slip velocity for $N=0.3$ are practically independent of mesh refinement, it is observed that for $N=13$, convergence is only achieved for the slip velocity, while the slip length is approaching convergence but is not yet fully mesh independent. This outcome is attributed to the observation that, for higher viscosity ratios, the velocity profile in proximity to the interface exhibits a less linear character compared to the behaviour observed for lower values of $N$. Consequently, while the velocity converges with reduced resolutions, the velocity gradient (utilised to calculate the slip length) necessitates higher refinement levels, resulting in a solution that is more contingent on the mesh size. Given the increase in computational cost associated with higher mesh refinement, it was decided to maintain a refinement level of 8, which is sufficient to ensure the mesh independence of the velocity field and the slip length for the majority of cases, yielding only minor errors for $N=13$. As these results were utilised in a comparison with experimental data, for which the slip values were considerably smaller than those computed from the simulations, these errors do not affect the results or the conclusions. It is also noteworthy that this convergence occurred at the interface, while the majority of results are presented for $y=0$, where the mesh convergence is reached for lower refinement.

%

\section{Experimental estimation of the average slip length}\label{appE}

The experimental setup used in the study does not allow to measure accurately the flow rate. Therefore average values of measurements are used to characterise the net effect of the LIS on the upper flow.
In the following, we compare two approaches to calculate the average slip length and identify the most suitable for LISs with a non-flat, patterned substratum. 

\begin{figure}
\centering
    \tikzset{every picture/.style={line width=0.75pt}} 
    \raisebox{-1\height}{\begin{tikzpicture}[x=0.75pt,y=0.75pt,yscale=-1,xscale=1]
    \node [inner sep=0pt,below right,xshift=0.0\textwidth] 
                {\includegraphics[width=0.43\textwidth]{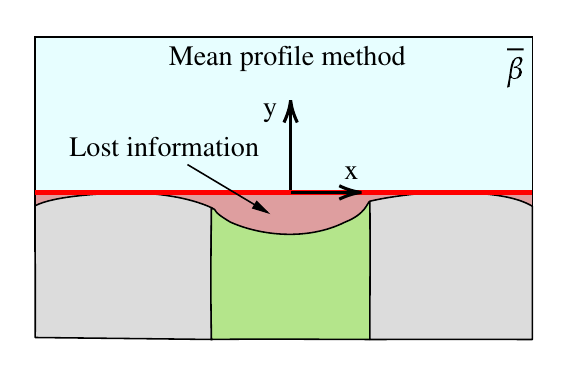}};
    \draw (-10,15) node [anchor=north west][inner sep=0.75pt]   [align=left] {($a$)};
    \end{tikzpicture}}
     \tikzset{every picture/.style={line width=0.75pt}} 
    \raisebox{-1\height}{\begin{tikzpicture}[x=0.75pt,y=0.75pt,yscale=-1,xscale=1]
    \node [inner sep=0pt,below right,xshift=0.0\textwidth] 
                {\includegraphics[width=0.43\textwidth, trim={0 0 0 0},clip]{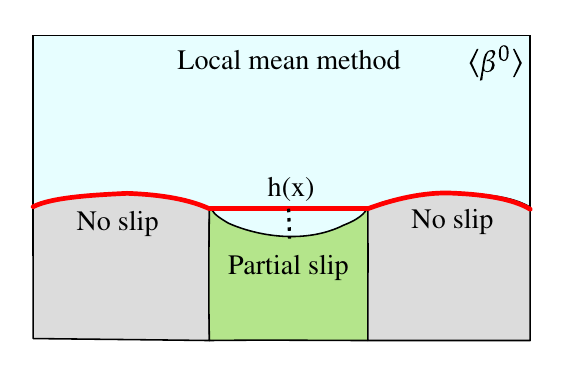}};
    \draw (-10,15) node [anchor=north west][inner sep=0.75pt]   [align=left] {($b$)};
    \end{tikzpicture}} 
    \caption{Illustration of two methods for the calculation of $\mathrm{\beta_{eff}}$ from measurements. ($a$) In the mean profile method, the horizontal plane tangent to the ridges' top is considered. ($b$) In the local mean method, the boundary is composed of the profile of the ridges and the height function $h(z)$ at the liquid interface. The top right corners remind of the variable obtained with each method.}
    \label{fig:Beta_methods}
\end{figure}

\begin{figure}
\centering
    \tikzset{every picture/.style={line width=0.75pt}} 
    \raisebox{-1\height}{\begin{tikzpicture}[x=0.75pt,y=0.75pt,yscale=-1,xscale=1]
    \node [inner sep=0pt,below right,xshift=0.0\textwidth] 
                {\includegraphics[width=0.6\textwidth]{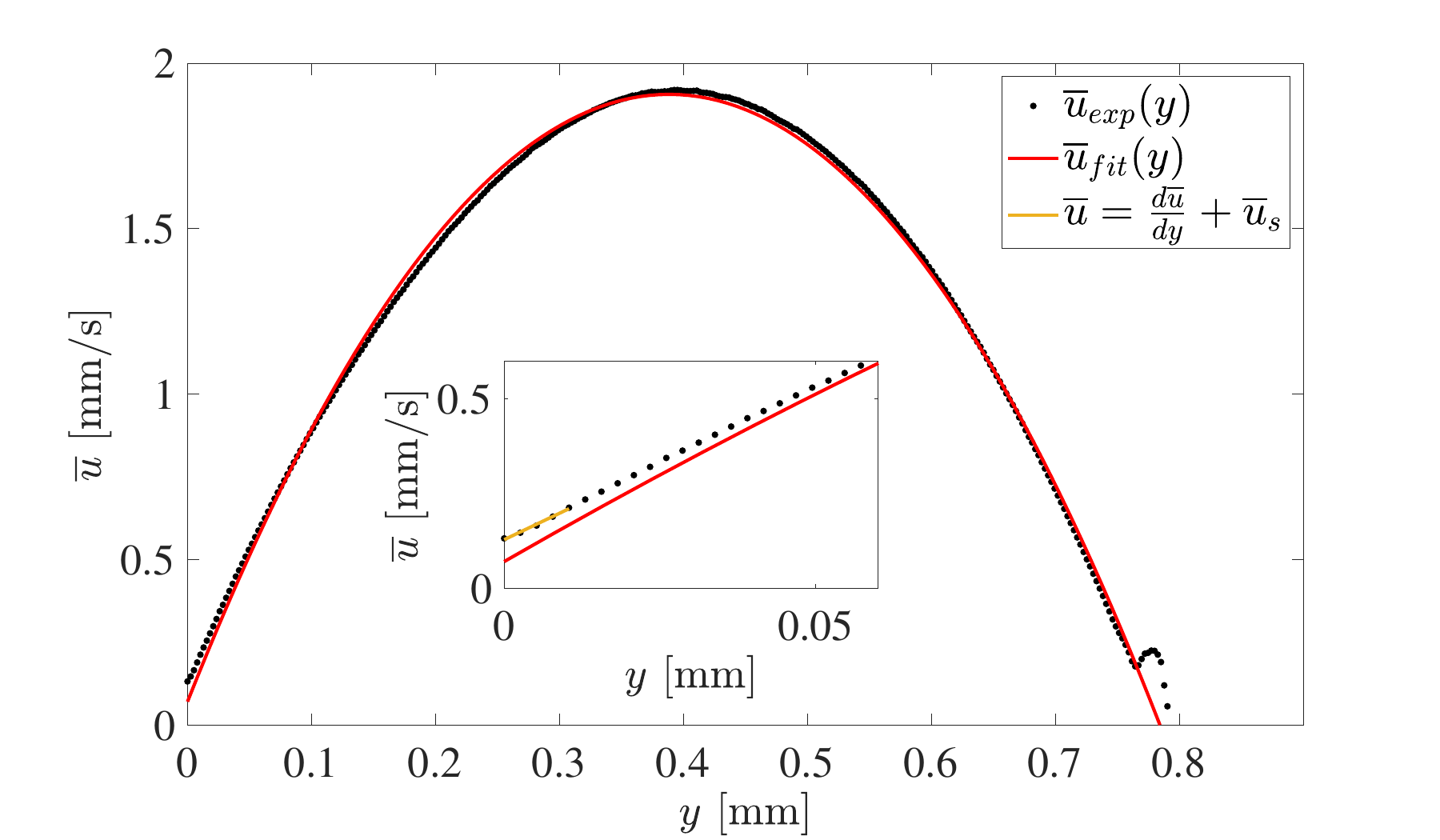}};
    \draw (10,10) node [anchor=north west][inner sep=0.75pt]   [align=left] {($a$)};
    \end{tikzpicture}}
     \tikzset{every picture/.style={line width=0.75pt}} 
    \raisebox{-1\height}{\begin{tikzpicture}[x=0.75pt,y=0.75pt,yscale=-1,xscale=1]
    \node [inner sep=0pt,below right,xshift=0.0\textwidth] 
                {\includegraphics[width=0.3\textwidth, trim={0 0 0 0},clip]{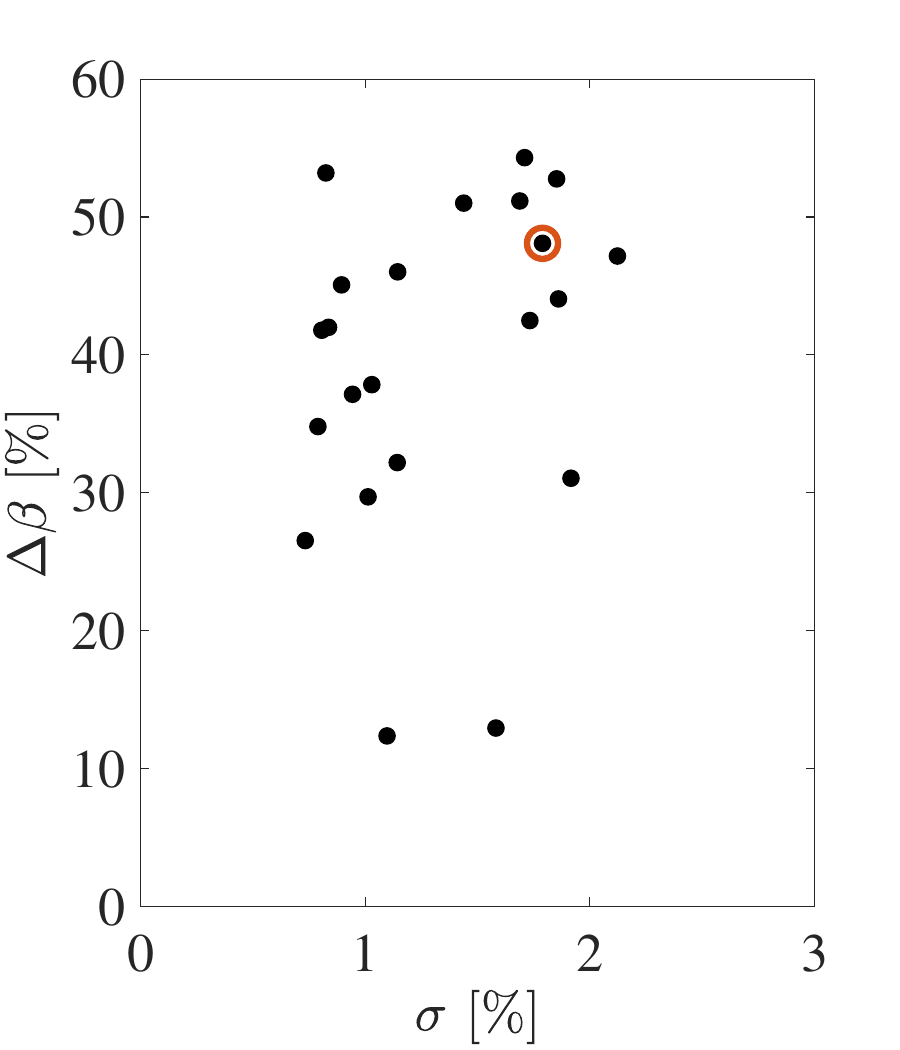}};
    \draw (-10,10) node [anchor=north west][inner sep=0.75pt]   [align=left] {($b$)};
    \end{tikzpicture}} 
    \caption{Sensitivity analysis for the mean profile method. ($a$) Mean velocity profile for Case A106N13 fitted to Equation \ref{eq:fit} (red line). The insert shows an enlargement of the near wall area where the linear fit of the mean profile is added. ($b$) The relative difference of the effective slip length calculated as a fitting parameter and according to Navier's definition with respect to the relative error of the fit. The red circle marks the case of panel ($a$).}
    \label{fig:sensitivity}
\end{figure}

The first approach referred to as the \textit{mean profile method}, has been employed in previous studies \citep{Bolognesi2014EvidenceMicrochannel,Schaffel2016LocalSurfaces}. In this method, the velocity profiles are averaged over the span of a pitch, an operation that implies considering as a reference the plane tangent to the crest of the solid pattern, as sketched in Figure \ref{fig:Beta_methods}$a$. The resulting profile is then fitted to a function that describes a pressure-driven flow with a partially slippery wall:
\begin{equation}
    \overline{u}_{\textit{fit}}(y)=4U_{m}\frac{(\beta_{\textit{fit}}+H)(H-y)[\beta_{\textit{fit}}(y+H)+y 
    H]}{H^2(2\beta_{\textit{fit}}+H)^2}
    \label{eq:fit}
\end{equation}
with the maximum velocity of the mean profile, $U_{m}$, the channel height $H$ and the effective slip length $\beta_{\textit{fit}}$ being the fitting parameters. An example of this analysis applied to the present data is shown in Figure \ref{fig:sensitivity}$a$, where the experimental mean profile, $\overline{u}_{\textit{exp}}(y)$, is fitted to Equation \eqref{eq:fit}. This method performs well if the ridge of the solid structure is flat, and the no-slip velocity regions can be fully captured during the averaging process. However, in the case of a non-flat solid boundary, information loss can occur where the reference plane deviates from the actual structure, as illustrated in the sketch of Figure \ref{fig:Beta_methods}$a$. As a result, the slip velocity of the mean profile is overestimated, leading to an overestimation of the effective slip length.
To quantify the accuracy of this method, we compared $\beta_{\textit{fit}}$ with the effective slip length obtained as $\overline{\beta}=\overline{u}_s/\frac{\partial \overline{u}}{\partial y}$.

The relative uncertainty of the fit is estimated as:
\begin{equation}
    \sigma=\text{rms}\left(\frac{u}{U_m},\frac{\overline{u}_\textit{fit}}{U_m}\right)
\end{equation}
and the relative difference between the effective slip lengths is given by:
\begin{equation}
   \Delta\beta=\frac{\overline{\beta}-\beta_{\textit{fit}}}{\overline{\beta}}.
\end{equation}
As shown in Figure \ref{fig:sensitivity}$b$, although the uncertainty of the fit is only \SI{1}{\percent}, the relative difference $\Delta\beta$ ranges from \SI{13}{\percent} in the best case to \SI{54}{\percent} in the worst case, highlighting the sensitivity of this method to the averaged velocity.

The second approach, named \textit{local mean method}, involves a reverse procedure compared to the previous one. In this case, the liquid-solid and liquid-liquid interfaces are treated separately, as depicted by the red profile in Figure \ref{fig:Beta_methods}$b$. The local slip length considered here is $\Expz$ described in Section \ref{sec:dns_exp}.
The average slip length is then determined by averaging these local values along the streamwise and spanwise directions over the span of a pitch and it is denoted as $\Expm$. 


\section{Effective slip length vs average slip length }\label{appF}
\begin{figure}
\centering
    \tikzset{every picture/.style={line width=0.75pt}} 
    \raisebox{-1\height}{\begin{tikzpicture}[x=0.75pt,y=0.75pt,yscale=-1,xscale=1]
    \node [inner sep=0pt,below right,xshift=0.0\textwidth] 
                {\includegraphics[width=0.46\textwidth]{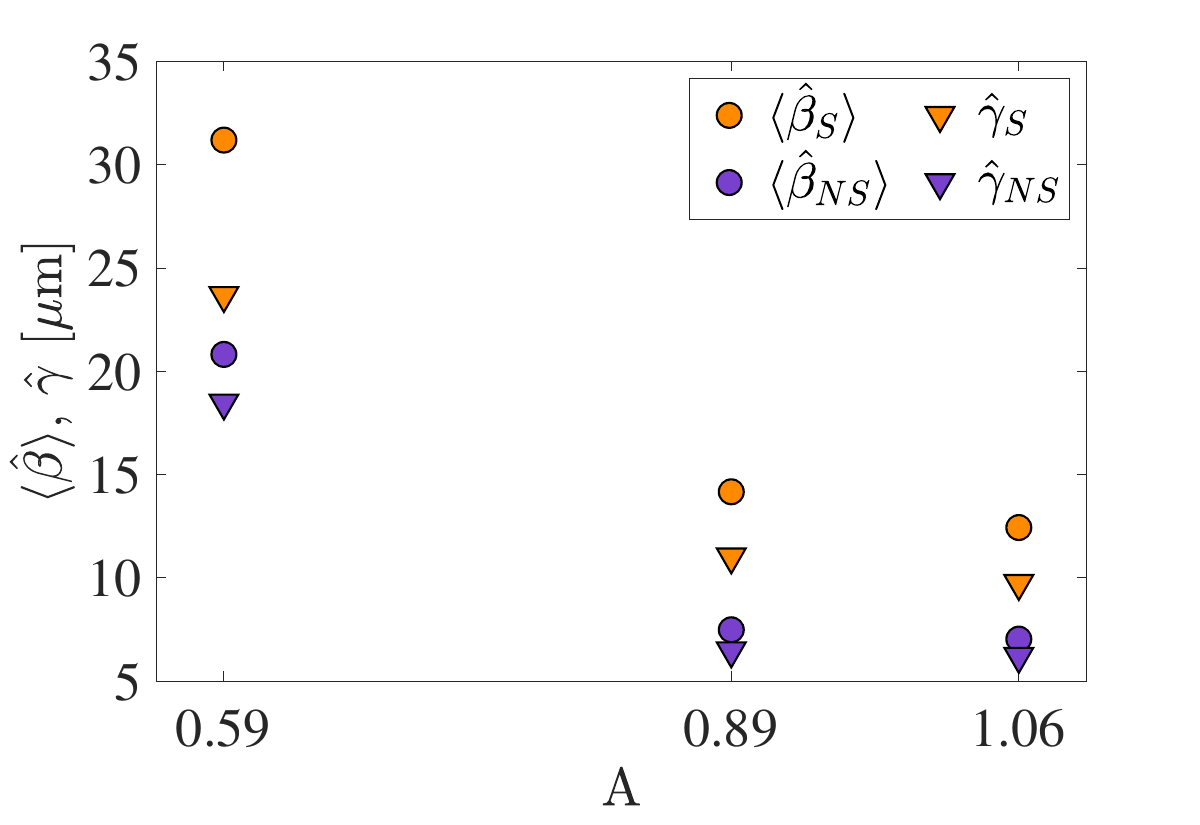}};
    \draw (-10,10) node [anchor=north west][inner sep=0.75pt]   [align=left] {($a$)};
    \end{tikzpicture}}
    %
    \tikzset{every picture/.style={line width=0.75pt}} 
    \raisebox{-1\height}{\begin{tikzpicture}[x=0.75pt,y=0.75pt,yscale=-1,xscale=1]
    \node [inner sep=0pt,below right,xshift=0.0\textwidth] 
                {\includegraphics[width=0.46\textwidth, trim={0 0 0 0},clip]{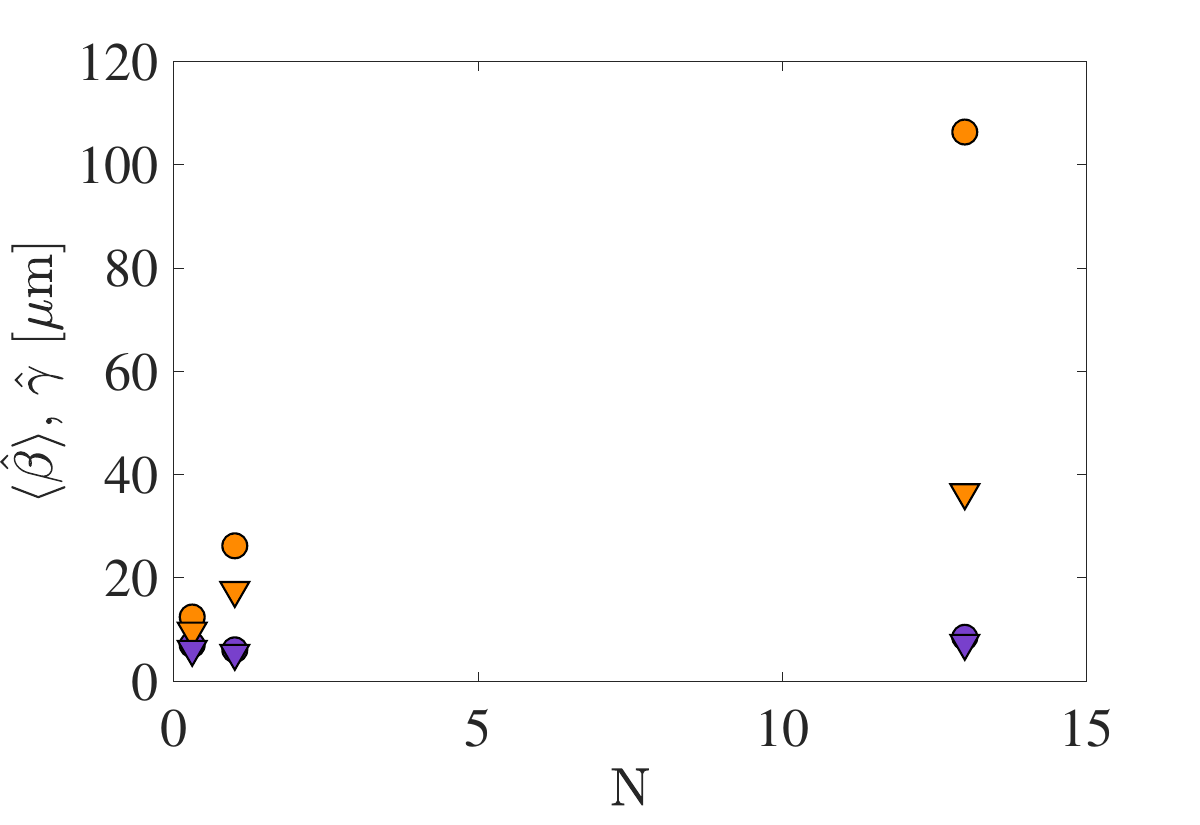}};
    \draw (-10,10) node [anchor=north west][inner sep=0.75pt]   [align=left] {($b$)};
    \end{tikzpicture}} 
    \caption{Effective slip length $\hat\gamma$ and average slip length $\langle\hat\beta\rangle$ derived from DNS data. The subscript "S" denotes \textit{slip} boundary condition and "NS" denotes \textit{no-slip velocity} boundary condition.}
    \label{fig:DNS-methods}
\end{figure}

 From the velocity field computed in the numerical simulation, the volumetric flow rate of the working fluid $Q$ can be calculated. 
 Using Equation \eqref{eq:dns_Q}
is possible to find the effective slip length for the surfactant-free, $\DNSeffs$, and the immobilized interface, $\DNSeffns$. This approach takes into account not only the surface slip but also the change in cross-section. On the other hand, considering Navier's definition, we derived $\DNSs$ and $\DNSns$ with equivalent boundary settings and we calculated their average values within a pitch: $\DNSms$ and $\DNSmns$.
Figure \ref{fig:DNS-methods}$a$ illustrates the results as a function of the groove aspect ratio and \ref{fig:DNS-methods}$b$ as a function of the viscosity ratio. Calculating the effective slip length using the second approach results in an overestimation of its magnitude. This discrepancy is more pronounced in the case of a surfactant-free interface and becomes increasingly significant with higher viscosity ratios and lower aspect ratios. The two approaches are equivalent in the case of a flat interface. However, the observed difference arises due to the meniscus curvature. The selection of the method to compute the effective slip length should therefore consider the specific interface geometry and the available experimental data used for comparison. It is worth noting that, for a no-slip velocity interface, the slip length remains nearly constant regardless of changes in the viscosity ratio.

\newpage

\bibliographystyle{jfm}
\bibliography{references}


\end{document}